\documentclass[letterpaper]{article} 
\usepackage{aaai25}  
\usepackage{times}  
\usepackage{helvet}  
\usepackage{courier}  
\usepackage[hyphens]{url}  
\usepackage{graphicx} 
\usepackage{natbib}  
\usepackage{caption} 
\usepackage{algorithm}
\usepackage{algorithmic}

\usepackage{newfloat}
\usepackage{listings}
\DeclareCaptionStyle{ruled}{labelfont=normalfont,labelsep=colon,strut=off} 
\floatstyle{ruled}
\newfloat{listing}{tb}{lst}{}
\floatname{listing}{Listing}
\usepackage{lipsum}
\usepackage{tikz}
\usepackage{booktabs}
\usepackage{makecell}
\usetikzlibrary{shapes.geometric, arrows.meta}
\usepackage[most]{tcolorbox}
\usepackage{multirow}
\usepackage{rotating}
\usepackage{graphicx}

\usepackage{xspace}
\newcommand{\hide}[1]{}
\newcommand{\xhdr}[1]{\vspace{1.7mm}\noindent{{\bf #1.}}} 

\newcommand{\eg}{\textit{e.g.,}\xspace}

\newcommand{\fig}{Figure\xspace}
\newcommand{\tab}{Table\xspace}
\newcommand{\sect}{\S}

\usepackage{xcolor}
\definecolor{ultramarine}{RGB}{0,32,96}
\definecolor{forestgreen}{RGB}{3,153,0}
\definecolor{mustard}{RGB}{234,170,0}

\newcommand{\answerYes}[1]{\textcolor{blue}{#1}} 
 
\newcommand{\answerNA}[1]{\textcolor{gray}{#1}}

\newcommand{\galen}[1]{}
\newcommand{\tim}[1]{}
\newcommand{\amy}[1]{}
\newcommand{\leon}[1]{}
\newcommand{\new}[1]{#1}

\title{Reddit Rules and Rulers: Quantifying the Link Between Rules\\and Perceptions of Governance Across Thousands of Communities}

\author{
Leon Leibmann, Galen Cassebeer Weld, Amy X. Zhang, Tim Althoff \\
\normalsize{{\normalfont Paul G. Allen School of Computer Science \& Engineering, University of Washington}} \\
\normalsize{\normalfont \{lleibm, gweld, axz, althoff\}@cs.washington.edu }
}

\begin{document}
\maketitle

\begin{abstract}
Rules are a critical component of the functioning of nearly every online community, yet it is challenging for community moderators to make data-driven decisions about what rules to set for their communities. The connection between a community's rules and how its membership feels about its governance is not well understood.
In this work, we conduct the largest-to-date analysis of rules on Reddit, collecting a set of 67,545 unique rules across 5,225 communities \new{which collectively account for more than 67\% of all content on Reddit. More than just a point-in-time study, our work measures how communities change their rules over a 5+ year period.} We develop a method to classify these rules using a taxonomy of 17 key attributes \new{extended from previous work}. We assess what types of rules are most prevalent, how rules are phrased, and how they vary across communities of different types.
Using a dataset of communities' discussions about their governance, we are the first to identify the rules most strongly associated with positive community perceptions of governance: rules addressing \textit{who} participates, how content is formatted and tagged, and rules about commercial activities. We conduct a longitudinal study to quantify the impact of adding new rules to communities, finding that after a rule is added, community perceptions of governance immediately improve, yet this effect diminishes after six months. Our results have important implications for platforms, moderators, and researchers. We make our classification model and rules datasets public to support future research on this topic.
\end{abstract}

\section{Introduction}\label{sec:intro}

Rules are critical to the safe and healthy functioning of online communities, and setting and enforcing rules is central part of their governance \cite{matias_2019_civic_labor, fiesler_redditrules_2018}. However, on many platforms, including Facebook Groups, Discord, and Reddit, community leaders have enormous leeway and minimal guidance in the rules they choose to set \cite{kraut_communities_2012}. Furthermore, measuring the relationship between rules and community outcomes \new{(such as bullying, misinformation, or community members' attitudes about governance)} is challenging due to the difficulty of quantifying both \cite{fiesler_redditrules_2018, weld_perceptions_2024}, as well as the abundance of confounding factors. As such, it's challenging for community moderators to know what rules would result in the best outcomes for their communities.

Studying rules in online communities poses several key difficulties. Making sense of the dozens of thousands of unique rules that are posted by thousands of varied communities requires a robust system for classifying rules. \new{Furthermore, communities change their rules as they grow and in response to events, meaning that point-in-time assessments of rules fail to capture temporal dynamics \cite{Reddy2023EvolutionOR}}. Previous work has examined rules across many communities \cite{fiesler_redditrules_2018, Hwang2022RulesAR, fang_2023_shaping}, and how rules are added as communities grow \cite{Reddy2023EvolutionOR, keegan_wikirules_2017}, yet how these rules relate to community outcomes is poorly understood. Community outcomes themselves are difficult to measure, and are often assessed with surveys \cite{weld_surveys_2021, koshy_2023_user_mod_alignment, almerekhi_2020_mod_harassment_modeling, seering_2022_twitch_moderator_recruiting, Kairam_2022_twitch_sovc}. Recently, methods have been proposed to automatically quantify different community outcomes \cite{weld_perceptions_2024, Bao_2021_prosocial}, yet this work does not directly address rules.

In this paper, we present the largest-to-date study of rules in online communities. We collect 67,545 unique rules from 5,225 communities constituting 67.58\% of all activity on Reddit, and identify how these rules were changed over a 5+ year study period, using data from the Wayback Machine (\sect\ref{sec:timelines}).
We develop a pipeline to classify these rules using a taxonomy of 17 different attributes, extended from previous work \cite{fiesler_redditrules_2018}.
Further, we use an existing method to quantify communities' public discussion of their governance \cite{weld_perceptions_2024} (\sect\ref{sec:sentiment}) along with causal inference methods (\sect\ref{sec:iptw}) and assess how rules are associated with community members' perceptions of their community's governance, \new{an important outcome}.

Our analyses address three key research questions:
\begin{itemize}
    \setlength{\itemindent}{6mm}
    \setlength{\itemsep}{0em}
    \item [\textbf{RQ1}] What rules do communities have on Reddit? (\sect\ref{sec:rq1})
    \item [\textbf{RQ2}] What rules are associated with positive community perceptions of governance? (\sect\ref{sec:rq2})
    \item [\textbf{RQ3}] What is the impact of adding new rules? (\sect\ref{sec:rq3})
\end{itemize}

We find that the rules set by communities vary widely, but important patterns emerge. Rules targeting post content are by far the most common, appearing in 89.74\% of communities, while rules about who can participate are relatively rare, appearing in only 20.89\% of communities (\sect\ref{sec:results_rules_number_size}). Communities are also more likely to phrase their rules in terms of what is \textit{prohibited} as opposed to permitted (\sect\ref{sec:results_pre_vs_re}). However, not all communities are the same: Discussion communities and communities for specific identity groups are $1.36\times$ more likely to have rules restricting who can participate  (\sect\ref{sec:results_difs_by_topic}).

Using IPTW causal inference methods \cite{Austin2015_IPTW_best_practice}, we adjust for confounding factors while assessing differences between communities \textit{with} vs. \textit{without} different rules. We find that rules about how content is formatted and who participates are associated with significantly more positive perceptions of governance (\sect\ref{sec:rq2}). Furthermore, we find that rules emphasizing what behavior is \textit{permitted} as opposed to \textit{prohibited} are also associated with more positive perceptions of governance. Finally, we conduct a longitudinal study of what happens in communities after new rules are added (\sect\ref{sec:rq3}), and find that rule additions are associated with an immediate improvement in communities' perceptions of their governance, although this effect typically diminishes after approximately six months.

We discuss important implications of our results for researchers, platforms, and community moderators, and highlight key opportunities for future work (\sect\ref{sec:discussion}).
We make our rule classification codebook, model, and rules data public\footnote{ \texttt{bdata.cs.washington.edu/mod-perceptions}} to support additional research on this important topic.
\section{Related Work}\label{sec:related}

\xhdr{Rules On and Off Reddit}
Rules have been previously studied on Reddit, including analyses categorizing rules that are present \cite{fiesler_redditrules_2018, fang_2023_shaping}, and analyses of how rules have changed \cite{Reddy2023EvolutionOR, lloyd_2024_airules, fang_2023_shaping}. However, our work is the first to directly connect rules to community outcomes by examining how community members perceive their governance (\sect\ref{sec:sentiment}).
Studies that assess rules' changes over time have used the Wayback Machine \cite{fang_2023_shaping, Reddy2023EvolutionOR} or manual scraping \cite{lloyd_2024_airules} to reconstruct timelines of communities' rules. In contrast to these studies, ours covers both a much longer time period (5+ years vs. 3 and 1.5) and a much larger set of communities (5,225 vs. 967 and 467). \citet{lloyd_2024_airules} includes rules from a larger set of communities than we do, yet they only measure if rules address AI generated content. In contrast, we assess 17 different attributes of rules.
Two taxonomies have been proposed for classifying Reddit rules \cite{fiesler_redditrules_2018, fang_2023_shaping}. We extend these taxonomies in our work, contributing a simpler taxonomy that captures a broader set of rule attributes (\sect\ref{sec:codebook}).


Off of Reddit, rules have been studied on peer production communities such as Wikipedia \& Wikia \cite{Hwang2022RulesAR, keegan_wikirules_2017, kittur_2010_wikia}, the fediverse \cite{Nicholson2023MastodonRC}, and gaming communities \cite{Frey2018EmergenceOI, Frey2022GoverningOG}.
While these are important other contexts, our work empirically evaluates rules specific to online social communities. 

\xhdr{Content Moderation and Governance}
Rules are also an important component of content moderation, which has been studied extensively on Reddit \cite{Proferes_2021_reddit_research_overview, li_2022_modlogs, matias_2019_civic_labor}. Some of this work has focused on specific aspects of content moderation, including banning users \cite{Cima2024TheGB, Thomas_2021_bans_behavior}, content removal \cite{jhaver_suspectremoved_2019, srinivasan_2019_content_removal_cmv, Ribeiro2020_migration} and removal reasons \cite{jhaver_2019_removal_reasons}. Other work has developed systems for participatory rule making \cite{Zhang2020PolicyKitBG} or has studied governance and norms through the lens of removed content \cite{eshwar_hiddenrules_2018}. Empirical analyses of removed content almost always do not enable the assessment of specific rules and their impact.  In contrast, in this work we examine rules directly, along with their connection with perceptions of community governance.

\xhdr{Community Values and Outcomes}
Understanding what it means to make communities `better' and what values are held by community members is a challenging problem \cite{Weld_2024_taxonomy}. Previous work has shown that values vary dramatically between different communities, and that what's good for one community is not necessarily good for another \cite{weld_surveys_2021, prinster_2024_archetypes}.
Other research has taken an ecological approach to understand how related communities overlap in their membership \cite{zhu_2014_wikia_membership} and rely upon one another \cite{hwang_2021_small_communities, TeBlunthuis_2022_similar_communities}.
Surveys are often used to assess community governance and community health \cite{koshy_2023_user_mod_alignment, almerekhi_2020_mod_harassment_modeling, seering_2022_twitch_moderator_recruiting, Kairam_2022_twitch_sovc}. 
As surveys do not scale well, some methods have been proposed to automatically quantify aspects of community outcomes \cite{Bao_2021_prosocial}. In this work we use an existing dataset of community members' stated perceptions of governance in their communities \cite{weld_perceptions_2024} to empirically assess rules at a scale not feasible with surveys or qualitative methods.

\section{Methods \& Data Collection}\label{sec:data}
To understand rules on Reddit, we first must know what communities had which rules when. We collect timelines of how communities' rules change using data from the Wayback Machine (\sect\ref{sec:timelines}), and classify those rules according to their tone, target, and topic using a GPT-4o-based retrieval-augmented few-shot multilabel classifier that achieves near-human performance (\sect\ref{sec:codebook}). Our work is the first to assess associations between what rules a community has, and how a community perceives its governance. To measure communities' perceptions of their governance, we use a classification pipeline developed by~\citet{weld_perceptions_2024} to identify posts and comments discussing community governance (\sect\ref{sec:sentiment}). \new{Finally, after identifying community size and topic as key factors associated with a community's rules in \sect\ref{sec:rq1}, we use causal inference methods (\sect\ref{sec:iptw}) to adjust for these confounding factors in subsequent analyses to further identify how rules impact a community's perceptions of its governance}. Throughout our analyses, in addition to the data collected using methods described below, we also use metadata (\eg for community size) computed from Pushshift data \cite{baumgartner_pushshift_2020} and community topics classified by \citet{weld_perceptions_2024}.

\begin{table*}[tb]
   \centering
   \includegraphics[width=1\textwidth]{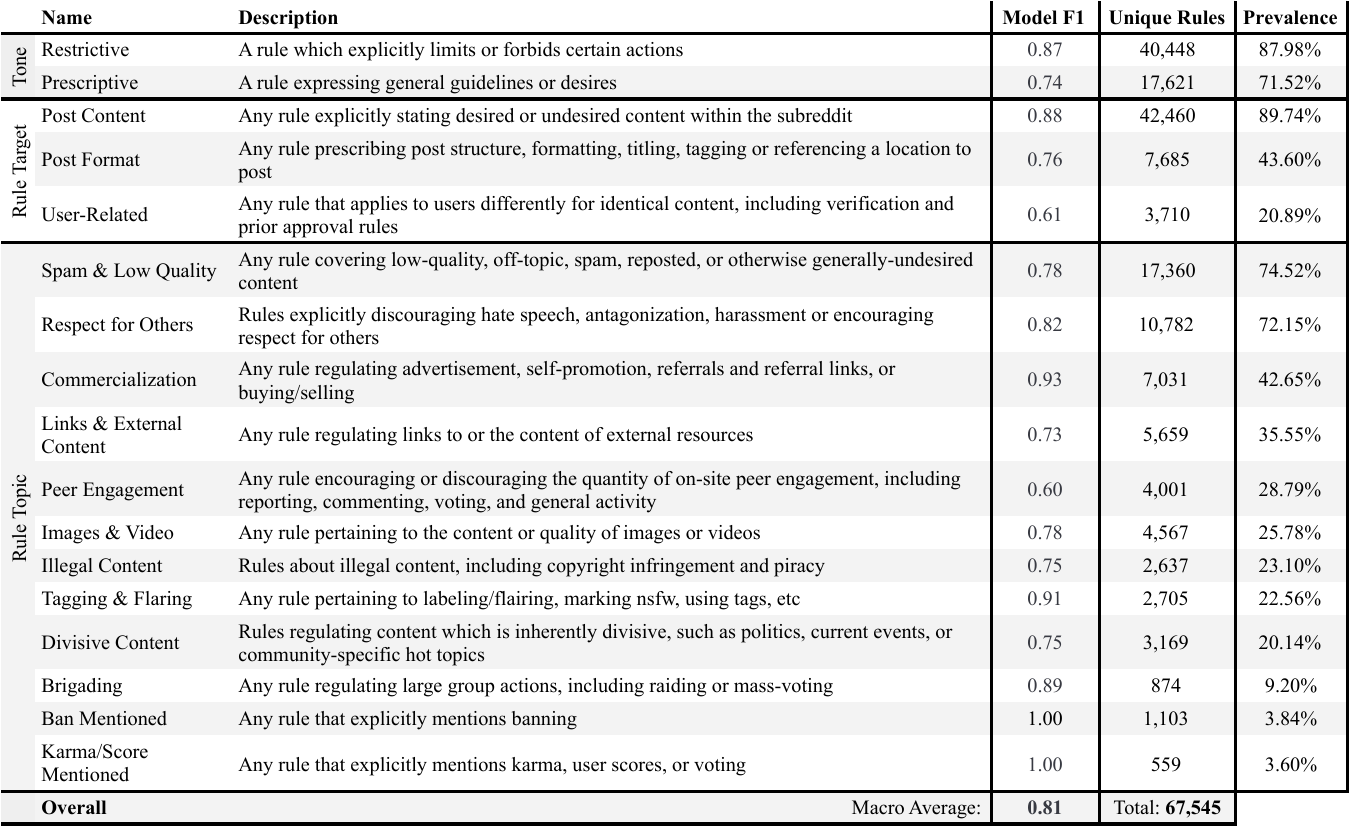}
   \vspace{-5mm}
 \caption{Our taxonomy of rules consists of 17 rule attributes across three different aspects of rules: tone, target, and topic. This breakdown of rule labels shows the performance of our retrieval-augmented \textbf{GPT-4o} classification model (Macro F1 = 0.81) and how likely to be encountered each rule type is (prevalence) across our 5+ year study period. Post Content is the most prevalent rule target, while Spam \& Low Quality and Respect for Others are the most prevalent rule topics.
 }
 \label{tab:codebook}
\end{table*}

\begin{table*}[tb!]
\centering
\small
\resizebox{.6\textwidth}{!}
{
\begin{tabular}{ll|l|l}
                             & \textbf{Our Taxonomy}                & \textbf{\citet{fiesler_redditrules_2018}}  & \textbf{\citet{fang_2023_shaping}}  \\ \hline
\multirow{2}{*}{\rotatebox[origin=c]{90}{Tone}}   & Restrictive                          & Restrictive                                &                                    \\ \cline{2-4} 
                             & Prescriptive                         & Prescriptive                               &                                    \\ \hline
\multirow{3}{*}{\rotatebox[origin=tr]{90}{Rule Target \hspace{0.1em}}} & Post Content                         & Content/Behavior                           &                                    \\ \cline{2-4} 
                             & Post Format                          & Format                                     & \makecell[l]{Requires a minimum text\\Requires template for post\\Limits use of title vs. description text} \\ \cline{2-4} 
                             & User-Related                         &                                      &\\ \hline
\multirow{21}{*}{\rotatebox[origin=c]{90}{Rule Topic}} & \multirow{4}{*}{\parbox{2cm}{Spam \& Low Quality}} & Spam                                 &No jokes\\
                             &                                      & Low-Quality Content                  &Posts must be high-quality\\
                             &                                      & Reposting                            &Content must be original\\
                             &                                      & Off-topic                            &Posts must be on-topic\\ \cline{2-4} 
                             & \multirow{3}{*}{Respect for Others}  & Trolling                             &\multirow{3}{*}{\makecell[l]{Enforces respect for others\\No promotion of bad behavior}}\\
                             &                                      & Hate Speech                          &\\
                             &                                      & Harassment                           &\\ \cline{2-4} 
                             & \multirow{2}{*}{Commercialization}   & Advertising \&     &\multirow{2}{*}{No promotional content}\\
                             &                                      & Commercialization                    &\\ \cline{2-4} 
                             & Links \& External                    & Links \& Outside Content             &Posts must be verifiable\\ \cline{2-4} 
                             & Peer Engagement                      & Personality                          &Enable peer feedback\\ \cline{2-4} 
                             & Images \& Video                      & Images                               &\\ \cline{2-4} 
                             & \multirow{3}{*}{Illegal Content}     & Copyright/Piracy                     &\\
                             &                                      & Doxxing/Personal Info                &\\
                             &                                      & Reddiquette/Sitewide                 &\\ \cline{2-4} 
                             & \multirow{2}{*}{Tagging \& Flaring}  & NSFW                                 &\\
                             &                                      & Spoilers                             &\\ \cline{2-4} 
                             & Divisive Content                     & Politics                             &\makecell[l]{Discourages divisiveness\\Avoid distressing material}\\ \cline{2-4} 
                             & \multirow{2}{*}{Brigading}           & Personal Army                        &\\
                             &                                      & Voting                               &\\ \cline{2-4} 
                             & Ban Mentioned                        & Consequences/Moderation/Enforcement             &\\ \cline{2-4} 
                             & Karma/Score                &                                      &\\ \hline
\end{tabular}
}
\caption{A comparison of our taxonomy with those from \citet{fiesler_redditrules_2018} and \citet{fang_2023_shaping}. Our taxonomy captures elements of rules not captured by previous taxonomies (\eg User-related rules and Karma/Score), while combining together some categories from previous taxonomies (\eg \citet{fiesler_redditrules_2018} provides four categories for Spam, Low-Quality Content, Reposting, and Off-topic, while we combine these into a single Spam \& Low Quality attribute).}
\label{tab:fiesler_comparison}
\end{table*}

\subsection{Computing Timelines of How Rules Change}\label{sec:timelines}
To collect timelines of how communities' rules change over time, we utilize the Wayback Machine \cite{wayback_machine}. Communities on Reddit post their rules on the sidebar of their homepages, where they are archived by the Wayback Machine\footnote{
\new{In 2019, Reddit introduced additional Rules pages which supplement the Rules posted in the sidebar. As these pages largely duplicate the information presented in the sidebar, are not archived by the Wayback Machine, and not available for the entire study period, we focus on sidebar content in this work.}
}.
\new{
We sought to collect rules timelines for as large and varied of a set of communities as possible.
We downloaded snapshots for approximately every English-language community with at least one snapshot every six months, for a set of 67,545 unique rules from 5,225 communities that collectively account for 67.58\% of all activity on Reddit during our 5+ year study period from April 2018 to December 2023. As the Wayback Machine archives more popular communities more often, most communities in our study are snapshotted more frequently than than every six months: 2,316 communities have at least monthly snapshots, and these communities account for 83.36\% of the community content included our study.
}




Next, we computed differences between snapshots to identify \textbf{Rules Periods}, blocks of time where a given community has an constant, unchanged set of rules. Each rule period consists of a set of one or more rules present during that period, along with time that the period began and ended. 
\new{Due to temporal gaps between snapshots, we have limited temporal resolution. To understand how this may impact our analyses}, using the time between subsequent snapshots, we compute the temporal uncertainty around when exactly each rules period began and ended. The average start/end uncertainty is $\pm17$ days, relatively small compared to the average period length of $378$ days. Therefore, we believe the limited temporal resolution for some communities has a negligible impact on our analyses.

To understand how common different rules are, we compute \textbf{Rule Prevalence}, which is the fraction of rules periods containing at least one rule of a given type, weighted by the duration of the rule period. Thus, rule prevalence measures the average likelihood of encountering a rule across all communities\footnote{For simplicity, we refer to this throughout the manuscript as `the fraction of communities with a rule of type $x$. However, as communities occasionally change their rules, a more precise interpretation would be `the fraction of communities with a rule of type $x$ when sampled uniformly across all communities and time across our study period. }.


\subsection{Automatically Classifying Rules' Tone, Target, and Topic}\label{sec:codebook}
On Reddit, rules are arbitrary text strings that are highly varied across communities. To perform our analyses of 67,545 unique rules, we developed a system to automatically classify rules' tone, target, and topic.

\xhdr{Taxonomy Development}
Starting with a random sample of 100 rules as a development set, a team of two researchers used a grounded theory approach to iteratively categorize the rules, using an inductive coding method~\cite{MacQueen1998CodebookDF}. 
The researchers independently clustered similar rules, then came together to resolve differences and reach consensus. Tentative clusters were assigned names and definitions to produce a working taxonomy, then the development set was recategorized, creating and removing categories by consensus. After two rounds of iteration, the process converged. At this point, the researchers independently labeled a separate test set of 200 randomly sampled rules, which was used to evaluate human IRR (Fleiss' kappa between 0.61 `substantial' and 0.91 `almost perfect' was achieved for all categories, with a Macro average of 0.83 `almost perfect') \cite{landis_cohenkappa_1977}.
The resulting taxonomy consists of 17 different rule attributes across three aspects of rules: tone, target, and topic. Table~\ref{tab:codebook} gives a description of each class. After resolving disagreements, our set of 200 rules labeled by two annotators was used as test set for the evaluation of our automated classification pipeline.

\textbf{Rule Tone} captures how the rule is phrased: prescriptive rules tell community members what \textit{to} do (\eg `Be nice!') while restrictive rules tell community members what \textit{not} to do (\eg `Don't be mean!').
\textbf{Rule Target} distinguishes between what type of interaction the rule is targeting: Rules can address the content of what is posted in a community (Post Content, \eg `no off-topic content allowed'), how that content is formatted (Post Format, \eg `Posts must begin with `ELI5''), or the users who post the content (User-Related, \eg `You must be approved by the mods to post'). Finally, the \textbf{Rule topic} classes focus on specific topics addressed by rules, such as Tagging \& Flaring, Spam, and Respect for Others.

\xhdr{Comparison with Previous Taxonomies}
Our taxonomy extends two previous taxonomies of rules \cite{fiesler_redditrules_2018, fang_2023_shaping}. Our taxonomy covers all codes from both previous taxonomies while adding two new attributes not directly covered (User-related rules and rules addressing Karma \& Score). 
Our taxonomy simplifies the \citet{fiesler_redditrules_2018} taxonomy, with with a slightly smaller set of attributes (17 vs. 24). To achieve this simplification, we merge several of the previous taxonomy's codes.
The \citet{fang_2023_shaping} taxonomy consists of 15 specific rule types. In contrast, our taxonomy covers a broader set of attributes (\tab\ref{tab:fiesler_comparison}). \new{
Compared to the rules taxonomy from \citet{fiesler_redditrules_2018} while adopting a hierarchical structure. Our 17 rule attributes are divided into three higher-level aspects of rules: tone, target, and topic. Compared to the other taxonomy, our taxonomy directly addresses rules targeting \textit{who} is permitted to participate, as well as rules about Karma and Score. We consolidate several codes from \citet{fiesler_redditrules_2018} into a smaller number of rule topics: Spam, Low-Quality Content, Reposting, and Off-topic become a single rule topic. We use a single topic for Respect for Others, which includes Trolling and Harassment from \citet{fiesler_redditrules_2018}. Likewise, we include a single topic for Illegal Content and for Brigading.
}

\new{
With regards to the taxonomy outlined by \citet{fang_2023_shaping}, our taxonomy adopts a broader perspective while aligning with two of their three overarching categories: Post Content and Post Format. We identify four categories (Images \& Video, Illegal Content, Tagging \& Flairing, Brigading) not present in the other taxonomy, and collapse the 3 format codes (Minimum Text, Templates, Title vs Description) proposed into a single Post Format label. Our taxonomy additionally identifies individual rule tone, \textit{how} a rule is phrased (prescriptive vs. restrictive), as well as explicit references to Bans and to Karma/Score.
}




\xhdr{Classification Pipeline} Given the many thousands of rules we need to label, human labeling is infeasible. As such, we developed a retrieval-augmented few-shot classification pipeline using on GPT-4o \cite{openai_hello_2024}. After the codebook was finalized, one annotator labeled an additional 200 rule training set used to provide few shot examples, retrieved using embedding-similarity provided by Nomic \cite{nussbaum_nomicembed_2024} as examples for the LLM (6-shot classification). This pipeline approaches human performance, achieving near-human performance with a Cohen's Kappa of 0.71 and Macro-F1 of 0.74 on our test set (Table~\ref{tab:model_performance}). Our GPT-4o model also exceeds the performance of all other models we evaluated, including other top general purpose large-language models: GPT-4 \cite{openai_gpt4technicalreport_2024}, Mistral \cite{mistralai_mistral-7b-instruct-v03_2024}, and Llama-3 \cite{dubey_llama3herd_2024}, as well as a fine-tuned version of RoBERTa \cite{liu_roberta_2019}.
\new{
Previous studies classifying rules on Reddit have used logistic regression classification methods \cite{fiesler_redditrules_2018, fang_2023_shaping}. However, these studies both predate the introduction of LLMs. We instead use a few-shot LLM-based pipeline, which results in higher performance with much less training data.
}

To further reduce the cost of labeling, we assigned the same label to rules that were nearly identical (differing only in capitalization or having an edit distance of two or fewer characters). This edit distance threshold was selected such that the addition of `no\ ' would cause the rules to be labeled separately. This reduced the number of rules to classify to 51,404. Labeling these rules using our pipeline took six hours at a cost of \$491.39 worth of OpenAI API credits.

\begin{table}[tbh]
    \centering
    \scriptsize

\begin{tabular}{l|rr}
                                & {Kappa} & {Macro F1} \\ \hline
Human Expert Performance    & {0.83}      & {0.80}      \\ \hline
\textbf{RoBERTa} \cite{liu_roberta_2019} on 300 human labeled    & {0.18}      & {0.23}      \\
\textbf{RoBERTa} on 3k GPT-4o RAG labeled & {0.41}   & {0.52}      \\
\textbf{Mistral} \cite{mistralai_mistral-7b-instruct-v03_2024} retrieval-augmented                         & {-0.07}      & {0.39}      \\
\textbf{Llama-3} \cite{dubey_llama3herd_2024} retrieval-augmented                         & {-0.07}      & {0.40}      \\
\textbf{GPT-4} \cite{openai_gpt4technicalreport_2024} zero-shot                      & {0.66}      & {0.63}      \\
\textbf{GPT-4} \cite{openai_gpt4technicalreport_2024} retrieval-augmented            & {0.64}      & {0.70}      \\
\textbf{GPT-4o} \cite{openai_hello_2024} retrieval-augmented                      & \textbf{0.71}      & \textbf{0.74}      \\

\hline
\end{tabular}

    \vspace{-1mm}

    \caption{Our rule classification pipeline using GPT-4o exceeds the performance of all other models evaluated and approaches human performance. An identical six-shot prompting pipeline was used for Mistral \cite{mistralai_mistral-7b-instruct-v03_2024}, Llama-3 \cite{dubey_llama3herd_2024}, GPT-4 \cite{openai_gpt4technicalreport_2024}, and GPT-4o \cite{openai_hello_2024}. Complete prompts are given in Appendix~\ref{app:prompt}.}
    \label{tab:model_performance}
\end{table}

\begin{figure}
    \centering
    \includegraphics[width=\linewidth]{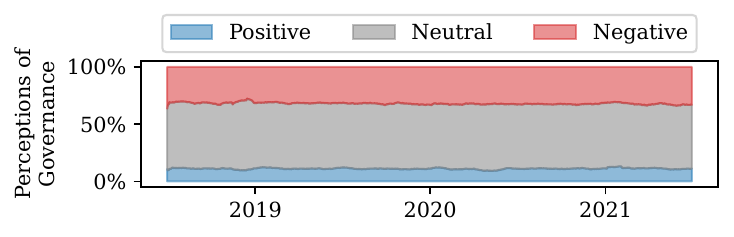}
    \vspace{-5mm}
    \caption{Across all communities on Reddit, community members' publicly expressed perceptions of their governance are approximately constant over time. Of posts and comments discussing governance, on average 11\% have positive sentiment, 57\% have neutral sentiment, and 32\% have negative sentiment.}
    \label{fig:perceptions_over_time}
\end{figure}

\subsection{Measuring Community Perceptions of Governance}\label{sec:sentiment}
Our work is the first to examine how online communities' rules are associated with how communities perceive their governance. To measure community members' perceptions of governance, we use the method developed by \citet{weld_perceptions_2024} to classify public posts and comments made on Reddit which discuss governance and moderators. This method consists of a three step pipeline to identify posts and comments discussing governance, filter out posts and comments addressing other communities, and classify the sentiment of each post and comment with regards to the governance using a using a QLoRA-tuned Llama2 model \cite{dettmers_QLoRA_2023}. 
\new{
To address the possibility that moderators' removal of comments critical of them might bias the validity of this pipeline, \citet[\sect3.4]{weld_perceptions_2024} examine a sample of removed comments, and conclude that this is highly unlikely: content discussing governance accounts for $<1\%$ of removed content.
}

We apply this method to all posts and comments made during our study period in communities for which we have rules timelines, collecting 3.8 million posts and comments discussing governance: 347 thousand posts and comments with positive sentiment, 1.9 million with neutral sentiment, and  1.5 million with negative sentiment.
For each community's rules periods, we compute the \textbf{Community's Perceptions of Governance} as the fraction of posts and comments discussing governance made in a given community during a given rules period having positive, neutral, and negative sentiment towards the community's governance.

\subsection{Adjusting for Confounding Factors}\label{sec:iptw}

In \sect\ref{sec:rq1}, we compute the prevalence of rules and identify two key factors that are associated with the rules set by each community: community topic and size. In subsequent sections, we 
examine which different types of rules are associated with more positive community perceptions of governance (\sect\ref{sec:rq2}), and what happens when new rules are added (\sect\ref{sec:rq3}). In each of these sections, we use a different method to adjust for these factors in order to better identify the potential impact of rules.

In \sect\ref{sec:rq2}, we use \textbf{Inverse Probability of Treatment Weighting (IPTW)}, a statistical method to adjust for confounding when comparing between different observed groups (\eg communities \textit{with} and \textit{without} rules of a given type). We use IPTW over other similar methods, such as stratification, due to its superior efficiency \cite{Austin2015_IPTW_best_practice}.

\new{
We begin with an intuitive example of how IPTW works.
Consider an analysis of rules about links and external content. Some communities have such rules, whereas others do not have such rules. We can compute the average perception of governance in both communities with and without links and external content rules, then take the difference between these averages to determine possible associations.
However, news-focused communities are more likely to have rules about links and external content (53.40\% vs. 33.79\% prevalence, see \fig\ref{fig:label_distribution}) and would thus be overrepresented in the set of communities with rules on links and underrepresented in the set of communities without these rules. To control for this selection effect, IPTW upweights underepresented communities and downweights overrepresented communities when computing average perceptions of governance across these groups. After weighting is performed, both groups have similar distributions of topics and size.
}

More formally, we compute IPTW weights using the probability of treatment (probability of having a rule of a given type), known as the propensity score, $P(Z|\mathbf{X})$ by applying a logistic regression model to community covariates $\mathbf{X}$ including the topic (one-hot encoded) and size of the community. For final analyses, each observation is weighted by the inverse of the probability of the treatment it received, such that the weight for observation $i$ is $w_i = \frac{Z_i}{P(Z_i=1|\mathbf{X}_i)} + \frac{1-Z_i}{P(Z_i=0|\mathbf{X}_i)}$, where $Z_i$ is the treatment received by observation $i$ (0 for control, 1 for treated), and $\mathbf{X}_i$ is a vector of the covariates.
\new{This weighting makes the distributions of covariates among the treatment and control groups more similar to the overall population.}

An important validity check for IPTW is to assess the balance of covariates for each treatment group after weighting~\cite{Austin2015_IPTW_best_practice}. 
Two groups are often considered `balanced' or `indistinguishable' if all covariates are within a standardized mean difference (SMD) of 0.25 standard deviations~\cite{althoff2016quantifying}.
Using the same procedure as in \citet{weld_perceptions_2024}, for each treatment/control group, we compute the difference between each covariate's weighted mean value and the \emph{reference distribution}, consisting of the entire population \new{of communities}. We compute the SMD by normalizing the difference in means by the standard deviation of the values of the reference distribution.
Across 238 condition-covariate-rule type pairs (17 experiments $\times$ 2 treatment/control conditions $\times$ 7 covariates), our method achieves balance in 235 cases (98.74\%). 

In the three cases where our method fails to achieve balance, no experiment has more than a single unbalanced covariate (out of seven), and no SMD exceeds 1.00. A complete list of covariates and SMDs for each experiment is given in Appendix~\ref{app:iptw_balance}, along with a comparison of weighted vs. unweighted results.
Appendix~\ref{app:iptw_balance} also shows the difference between IPTW results and those without any adjustment for confounding. We find that after performing IPTW, most confidence intervals do not overlap 0, suggesting that rules play an important role in communities' perceptions of governance, although community topic and size partially account for some of the differences between communities.

In \sect\ref{sec:rq3}, to measure what happens after rules are added, we use time series data and longitudinal analyses to quantify perceptions of governance before vs. after a rule is added. As such, because we are not comparing between different groups, IPTW is not feasible. Instead, the time-based format of our analyses inherently control for the potential community topic and size confounds, as these factors do not meaningfully change over a relatively short time period.

\begin{figure}[t]
   \centering
   \includegraphics[width=0.45\textwidth]{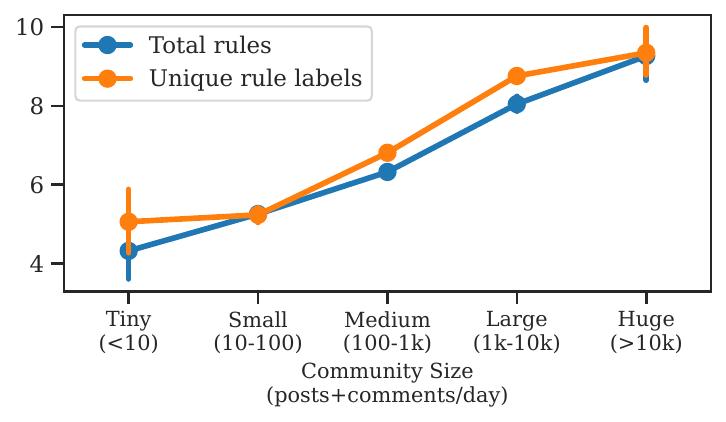}
   \vspace{-3mm}
 \caption{Larger communities have both more rules and more diverse rules. Tiny communities have on average 4.32 rules, while huge communities (the 0.77\% largest) have on average 9.26 rules. In this and subsequent figures, the bars shown represent bootstrapped 95\% confidence intervals.}
 \vspace{-2mm}
 \label{fig:average_rules}
\end{figure}

\subsection{Ethical Considerations}\label{sec:ethics}
We believe this work will have a positive broader impact by informing rule setting in online communities. We believe our work has minimal risk to participants' privacy as we only use public data, however, we take further steps to reduce potential harms and misuse potential of work: we do not publish moderator information, participant usernames or any identifiable information.
This study was approved by the University of Washington IRB under ID number STUDY00011457.

\section{What Rules Do Communities Have?}\label{sec:rq1}

\begin{figure}[htb]
   \centering
   \includegraphics[width=.9\linewidth]{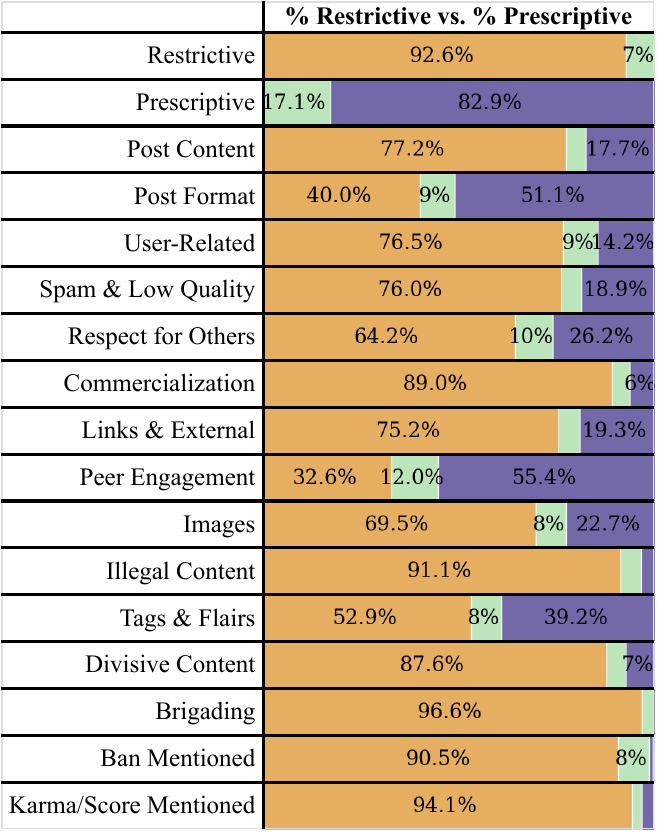}
   \includegraphics[trim=0 3mm 0 3mm,clip,width=.8\linewidth]{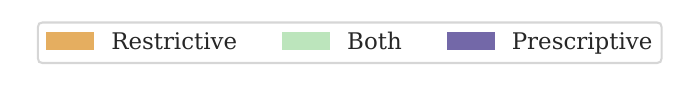}
   \vspace{1mm}
 \caption{Rules vary with regards to their tone. On the whole, restrictive rules are more commonly encountered than prescriptive rules on Reddit, although both are ubiquitous: 87\% of communities have at least one restrictive rule, while 70\% of communities have at least one prescriptive rule (prevalence). Rules addressing post format and peer engagement are both more likely to be prescriptive (`Be Nice') than restrictive (`Don't be mean'), while all other types of rules are more commonly phrased with restrictive tone.}
 \vspace{-2mm}
 \label{fig:label_and_likelihoods}
\end{figure}

\subsection{Number, Diversity, and Prevalence of Rules}\label{sec:results_rules_number_size}
\xhdr{Results}
We find that communities on Reddit vary dramatically in the number and types of rules they declare. 95\% of communities declare between two and 14 rules (median 7 rules), although in general, the larger the community, the more rules they declare (\fig\ref{fig:average_rules}). On average, communities with $<10$ daily posts and comments declare 4.32 rules, while huge communities (which account for 29.87\% of all posts and comments) have an average of 9.26 rules.

Some rules are much more frequently encountered than others. 89.74\% of communities have rules targeting post content (Table \ref{tab:codebook}), while rules targeting \textit{who} participates (User-Related rules) are present in only 20.89\% of communities. Rules about brigading, bans, and karma/score are the least common rule topics, occurring in only 9.20\%, 3.84\%, and 3.60\% of communities, respectively.

\xhdr{Implications}
As communities grow and age, they tend to add rules \cite{Reddy2023EvolutionOR}. Moderators on Reddit and other platforms often take a reactive approach \cite{matias_2019_civic_labor}, meaning that rules may be added only after an incident in a community makes it clear that such a rule is needed. The ubiquity of rules about spam, low quality content, and respect for others suggest that these are topics that most communities struggle with, and efforts by platforms and researchers to improve these aspects of communities are likely broadly useful.

\begin{figure*}[t]
   \centering
   \includegraphics[width=1\textwidth]{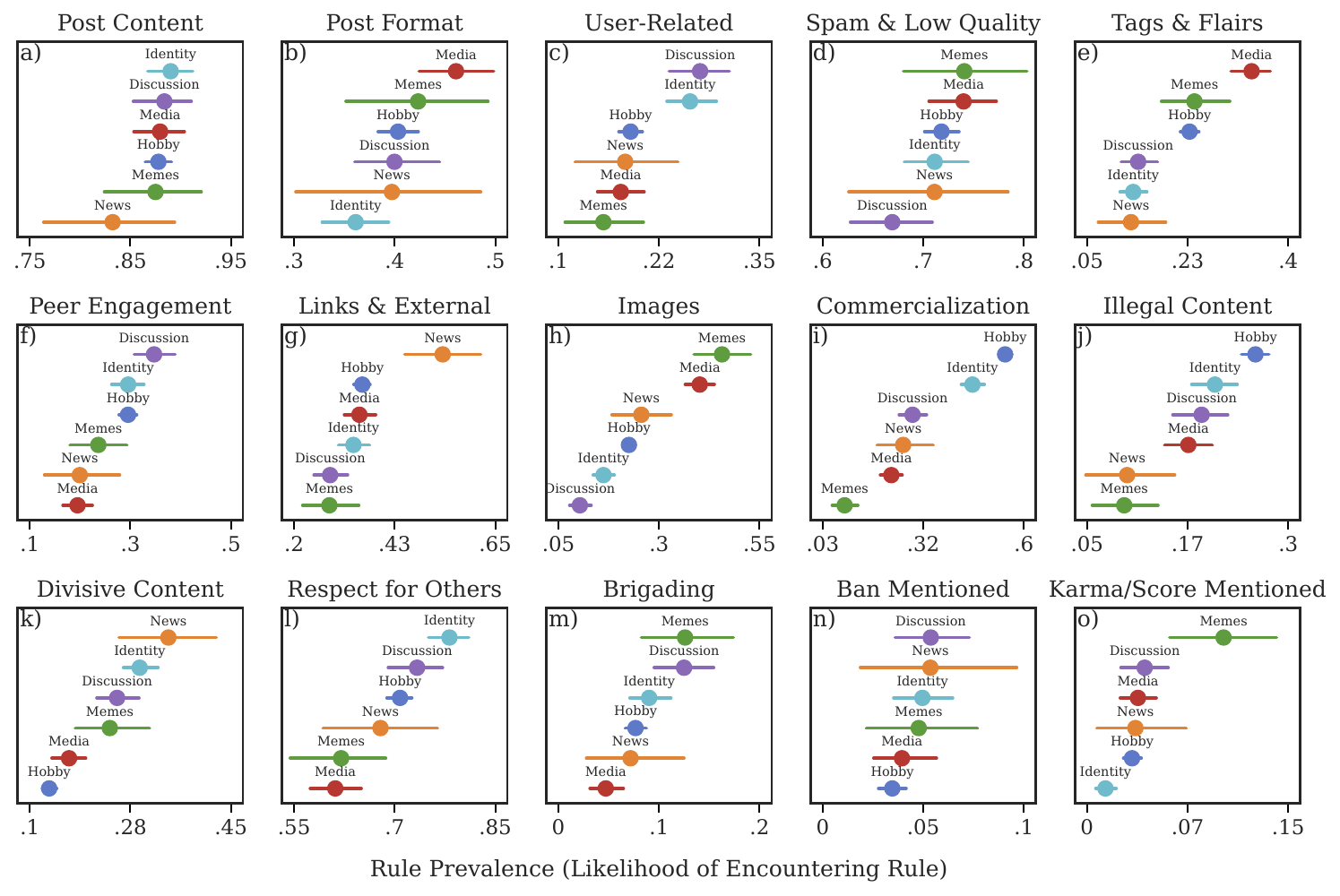}
   \vspace{-5mm}
 \caption{The ubiquity of different types of rules differs greatly based on community topic. Discussion and Identity communities are especially likely to have rules about who participates (c). News communities are \new{almost} twice as likely to have  rules about links and external content (g), on average, while Hobby \& Identity communities are more likely to have rules on Commercialization (i). Rules about Images are much more common in Meme and Media communities and very rare in Discussion (often text-based) communities (h).}
\vspace{-2mm}
 \label{fig:label_distribution}
\end{figure*}

\subsection{Prescriptive vs. Restrictive Rule Tone}\label{sec:results_pre_vs_re}
\xhdr{Results}
Rules can be phrased with prescriptive tone (`Be nice.') or restrictive tone (`Don't be mean.'), and occasionally phrased both prescriptively and restrictively (`Be  nice. Being mean is not allowed.'). We classify the tone of rules (\sect\ref{sec:codebook}), and find that on the whole, restrictive rules are more common than prescriptive rules, but both tones are widely used (\fig\ref{fig:label_and_likelihoods}), and 66.58\% of communities have both prescriptive and restrictive rules. The use of both tones simultaneously is relatively rare, with 94.22\% of rules being phrased only prescriptively or only restrictively. 

We find that certain types of rules are more frequently phrased restrictively, and others are more frequently phrased prescriptively. Rules about post format and peer engagement are both more frequently phrased using prescriptive tone (\eg `posts titles must be descriptive'). On the other hand, rules mentioning bans, karma/score, and brigading are almost often written using restrictive tone.

\begin{figure*}[tb]
   \centering
   \includegraphics[width=\textwidth]{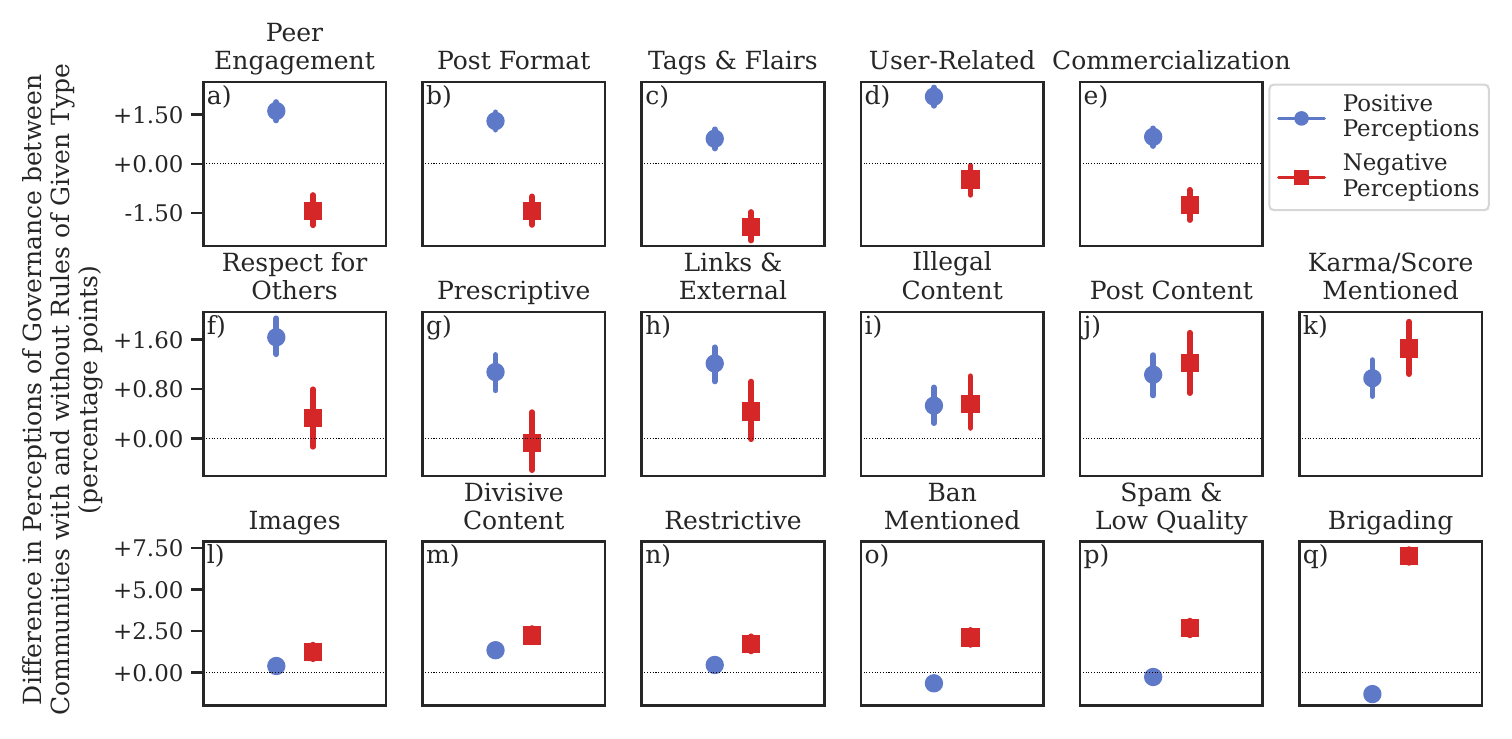}
   \vspace{-5mm}
 \caption{Perceptions of moderators vary between communities with and without different types of rules, even after adjusting for confounding factors (\sect\ref{sec:iptw}). Rules about Peer Engagement (a), Post Format (b), Tags \& Flairs (c) , and Commercialization (e) are all associated with higher positive perceptions and lower negative perceptions of moderators than communities without those rules. On the other hand, communities with rules about Illegal Content (i), Post Content (j), and Karma/Score (k) have more polarized perceptions of moderation, with positive \textit{and} negative perceptions \textit{both} higher than in other communities (which have higher neutral perceptions of moderation, not shown here). 
}
 \label{fig:sentiment_by_rule_type}
\end{figure*}

\xhdr{Implications}
Offline criminology literature has explored differences between telling people what they \textit{should} vs. \textit{shouldn't} do, which we examine in the online community context.
In \sect\ref{sec:rq2} we examine associations between rule tone and communities' perceptions of their governance, 
\new{
and find that communities with prescriptive rules tend to use more positive language to discuss their governance than those without prescriptive rules, whereas communities with restrictive rules tend to use more \textit{negative} language than those without restrictive rules (\fig\ref{fig:sentiment_by_rule_type}). 
}
Furthermore, our findings align with previous analyses of rules that are more likely to be phrased prescriptively \cite{fiesler_redditrules_2018}.

\subsection{Differences Between Communities with Different Topics}\label{sec:results_difs_by_topic}
\xhdr{Results}
Using an existing classification of communities based on their topic \cite{weld_perceptions_2024}, we examine how likely communities with different topics are to have different rules. While the prevalence of some rules varies substantially based on topic, other rules are applied relatively consistently.
Discussion communities and communities focused on Identity (such as those for LGBTQ groups) are $1.36\times$ more likely to have User-related rules than other communities (\fig\ref{fig:label_distribution}c). News-sharing communities are $1.58\times$ more likely to have rules about Links \& External Content (\fig\ref{fig:label_distribution}g), while Meme communities are $2.75\times$ more likely to have rules about Karma (\fig\ref{fig:label_distribution}o).
On the other hand, we do not find a significant difference between communities of different topics in their likelihood of having rules targeting post content (\fig\ref{fig:label_distribution}a), which are present in 89.74\% of communities. Similarly, only 3.84\% of communities have rules mentioning bans, and we do not find that this varies significantly based on the topic of the community (Fig.~\ref{fig:label_distribution}n).

\xhdr{Implications}
Our results suggest that certain types of rules, like those targeting the content of posts as well as specifically addressing Spam \& Low Quality content, are considered necessary by a majority of communities of every topic. Other types of rules are more necessary to communities with specific topics, such as rules about Commercialization, which are especially prevalent in Hobby communities. Communities should consider the specific needs of their community when determining what rules to post and enforce.

\section{What Rules Are Associated With Positive Community Perceptions of Governance?}\label{sec:rq2}

\xhdr{Method}
To understand what rules are associated with positive community perceptions of governance, we make comparisons between communities \textit{with} and \textit{without} rules of various types.
We measure perceptions of governance using an existing method to classify community members' discussion of their community's governance \cite{weld_perceptions_2024}. This dataset consists of posts and comments classified into positive, neutral, or negative sentiment with regards to community governance. For each community, we compute the fraction of posts and comments with each of these three sentiment classes (for more detail, see \sect\ref{sec:sentiment}) made during the entirety of each rules period.

When making comparisons between different communities, we seek to identify the differences in perceptions of governance that are associated with the presence or absence of different rules. Towards this goal, we adjust for two key confounding factors, community size and topic, using Inverse Probability of Treatment Weighting (IPTW) \cite{Austin2015_IPTW_best_practice}. For more details on IPTW, see \sect\ref{sec:iptw} and Appendix~\ref{app:iptw_balance}.

\xhdr{Results}
We find substantial differences in communities' perceptions of their governance between communities with and without various rules (\fig\ref{fig:sentiment_by_rule_type}).
\new{
Several rule types are associated with more positive and less negative perceptions of governance, particularly rules about Peer Engagement, Post Format, Tags \& Flairs, and Commercialization  (\fig\ref{fig:sentiment_by_rule_type}a-c,e).
}
We find that even though rules addressing \textit{who} participates are less common than rules addressing Post Content or Format, communities with User-Related rules have 1.83 percentage points more posts and comments expressing positive sentiment towards their governance after adjusting for confounding factors (\fig\ref{fig:sentiment_by_rule_type}d).
\new{
As only 11.12\% of posts and comments discussing governance in the average community  have positive sentiment, (\fig\ref{fig:perceptions_over_time}), a 1.83 percentage point increase would result in a 16.45\% increase in positive sentiment in a typical community. Put another way, after 1.83 pp increase, a huge\footnote{\label{size_defs}We use the same community sizes bins as in \fig\ref{sec:results_rules_number_size}: Large communities have between $10^3$ and $10^4$ posts \& comments per day, and Huge communities have $>10^4$.} community could expect almost 24 more comments per week expressing positive attitudes, while a large community could expect almost 6 more such comments per month.
}

However, some types of rules have more complex associations with perceptions of governance.
Rules about Illegal Content, Post Content, and Karma/Score are all associated with more \textit{polarized} perceptions of governance, having \textit{both} higher positive and higher negative perceptions of governance than communities without such rules (\fig\ref{fig:sentiment_by_rule_type}i,j,k). The correspondingly smaller fractions of posts and comments and comments with neutral sentiment towards governance are not shown.

\xhdr{Implications}
Our results show substantial differences in community perceptions of governance between communities with different rules. In particular, rules that promote positive interactions (Peer Engagement, User-Related, and Commercialization) and rules that structure contributions (Post Format, Tagging \& Flairing) are associated with the most positive perceptions of governance. Community moderators should consider how their rules address positive interactions and contribution structure.

We also find differences between rules with different tone. Rules with prescriptive tone (`do this') are associated with more positive perceptions of governance, whereas rules with restrictive tone (`don't do this'), are associated with more negative perceptions of governance (\fig\ref{fig:sentiment_by_rule_type}g,n).

\new{
We find that rules mentioning bans and brigading are associated with more negative and less positive perceptions of governance (\fig\ref{fig:sentiment_by_rule_type}o,q). These are both emotionally charged actions, which may contribute to the association with more negative perceptions of governance \cite{Thomas_2021_bans_behavior, Cima2024TheGB, Datta2018ExtractingIC}.
}

Overall, we find that many (13/15) rule types are associated with more positive perceptions of governance. Does adding new rules improve community perceptions of governance? We examine this question directly in \sect\ref{sec:rq3}.

\section{What Is the Impact of Adding New Rules?}\label{sec:rq3}
In \sect\ref{sec:rq2} we assessed what types of rules are most associated with favorable community perceptions of governance. But what happens when communities add new rules? We use our timelines of rules (\sect\ref{sec:timelines}) to conduct longitudinal analyses of how perceptions of governance (\sect\ref{sec:sentiment}) vary immediately after new rules are added to a community.\footnote{Rule additions are relatively rare events, with only 6,645 occurring during our 5 year study period. Although our data permit evaluation of different rule types, for these longitudinal analyses, confidence intervals are too large to be conclusive when analyzing the response to additions of rules of different types. As such, here we consider the addition of rules of \textit{any type}.
}

\xhdr{Method}
Starting with our set of rule periods, we compute \textbf{rule change} events for each community, where each rule change consists of two adjacent rule periods. We examine only rule changes where a rule that was not previously present is added, and, as this analysis examines a narrower time range and therefore requires higher temporal precision, we only include rule changes whose time of change is certain within $\pm1$ week precision. Across our 5+ year study period and 5,225 communities, this leaves a set of 6,645 rule change events matching these criteria. For each change event, we identify a 12 month long pre-change baseline window, and divide the 12 months post-change into $6\times$ two month long comparison windows, enabling analysis of how perceptions change over time\footnote{We experimented with a range of window widths and found that the exact window width does not make a qualitative difference in the results; thus we selected a two month width for the post-change comparison window to balance temporal resolution with statistical power. 
}.

For the pre-change baseline and each of the post-change comparison windows, we compute the perceptions of governance (fractions of posts+comments discussing governance having positive, neutral, and negative sentiment, for more details see \sect\ref{sec:sentiment}) for the relevant community. We then take the difference between the baseline and the comparison windows to see how perceptions of governance changed after the new rule was added. We average across all rule change events and bootstrap confidence intervals (as in all other analyses) to quantify our statistical power.

\begin{figure}[tb]
    \centering
    \includegraphics[width=\linewidth]{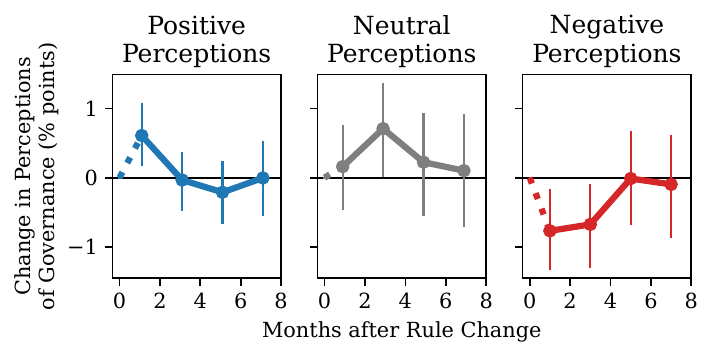}
   \vspace{-5mm}
    \caption{Immediately after a new rule is added, on average, positive perceptions of governance increases while negative perceptions of governance decrease. After approximately 6 months, this effect diminishes and community perceptions of governance are not significantly different than before the rule change.}
    \label{fig:rule_change_timeline}
\end{figure}

\xhdr{Results}
We find that, on average, immediately after a new rule is added, perceptions of governance become more favorable:
The fraction of posts and comments discussion governance positively \textit{increases} by 0.61 percentage points, while the fraction with negative sentiment \textit{decreases} by 0.77 percentage points (\fig\ref{fig:rule_change_timeline}).
\new{
For a typical huge community, this effect would be equivalent to almost 17 comments per week expressing negative perceptions of governance switching to expressing positive perceptions.
}
We also find that rule additions have a depolarizing effect on perceptions of governance, with the fraction of posts and comments discussing governance with neutral sentiment increasing by 0.71 percentage points in the 2-4 month period after a new rule is added.
Importantly, however, we find that the effects of new rule additions appear to `wear off' after approximately six months post-rule change, as the perceptions of governance returns to a state not significantly different from the pre-change baseline (\fig\ref{fig:rule_change_timeline}).

\xhdr{Implications}
Our results show the addition of new rules is associated with a small but statistically significant improvement in community members' publicly stated perceptions of the governance in their communities. This is consistent with the hypothesis that rules are often added as a response to an incident within a community that makes it clear such a rule is necessary. Our finding that the impact of a new rule on perceptions of governance at least partially diminishes after approximately six months is noteworthy. Part of the impact of a new rule comes not from the associated enforcement of that rule, but from the signal that the new rule provides that the moderator team is adjusting their strategy to the needs of the community. We speculate that this signaling aspect of rule changes contributes to the observed diminishing effect.

\section{Discussion \& Conclusion}\label{sec:discussion}
Rules, their communication, and their enforcement, are critical parts of the governance of nearly every online community.
Our results have important implications for platforms, community moderators, and researchers.

Our results inform what rules communities should have. We find that rules about Post Content, Spam \& Low Quality content, and Respect for Others are especially common (\sect\ref{sec:results_rules_number_size}). Platforms could consider providing rule `starter packs' for new communities based on common rules across the platform. We find that certain rules are more ubiquitous in communities of certain types, for example, Discussion and Identity communities are more likely to have User-related rules than other communities (\sect\ref{sec:results_difs_by_topic}). An interesting future research idea is a tool that recommends rules for moderators to consider implementing based on community size, topic, and other factors.
Our results also show that differences in rules explain some of the differences between different communities' perceptions of their governance (\sect\ref{sec:rq2}). Rules about Illegal Content, Post Content, and Karma/Score are associated with more polarized perceptions of governance. These rules maybe more polarizing because they relate to the boundaries of what behaviors and content are acceptable in a community, rather than how that content is presented. Rules that are polarizing are not necessarily bad for communities, but these are important impacts to consider.

Our results also have implications for how rules should be phrased. In many cases, moderators can choose to phrase rules to describe what community members \textit{should} do (prescriptive) vs. what they \textit{should not} do (restrictive). Our results suggest that certain rules are more easily phrased restrictively (\sect\ref{sec:results_pre_vs_re}), but that prescriptive rules are associated with more positive perceptions of governance, whereas restrictive rules are associated with more negative perceptions of governance (\sect\ref{sec:rq2}). Future research could evaluate a system that helps moderators consider alternate phrasings or presentations for their rules \cite{Matias_2019_Preventing}.


\new{
Our results show that new rules are associated with an immediate improvement in communities' perceptions of governance (\sect\ref{sec:rq3}). While we cannot know \textit{why} moderators added rules, any rule additions are a sign that moderators are engaging with their community, and our results are consistent with results from other studies showing that moderator engagement with community members is received positively \cite{weld_perceptions_2024}. As it is neither feasible nor necessarily a good idea for communities to continuously add new rules, our finding that the positive impact of adding a new rule diminishes after 6 months (\sect\ref{sec:rq3}) highlights the importance of other means for moderators  to engage with their communities, such as educating members about the rules \cite{jhaver_2019_removal_reasons, Matias_2019_Preventing} and soliciting feedback \cite{grimmelmann_2015_virtues_of_mod, kraut_communities_2012}.
}

\new{
For researchers, we make our dataset of rules, our taxonomy of rule attributes, and our classification methods public to support future research.
}

\subsection{Limitations}\label{sec:limits}

Our work establishes connections between the rules a community has and that community's perceptions of its governance, as measured using public discussion of governance and moderators (\sect\ref{sec:sentiment}). While we take great care to ensure the robustness of our results, including quantifying temporal uncertainty and reporting confidence intervals throughout, our work has several important limitations. The Wayback Machine, which we use to measure how rules change over time, has infrequent or absent snapshots for less popular and very small communities, thus potentially biasing our results. Additional research is needed to understand how very small communities and communities in languages other than English differ. We also only consider the rules that are posted by communities, not how these rules are enforced or how visible they are \cite{Matias_2019_Preventing}. Collecting information about rule enforcement is difficult, but not impossible \cite{eshwar_hiddenrules_2018}.

\new{
Rule modifications can occur for a variety of reasons.
While our work examines many thousands of rule changes, our methods are unable to track exactly why a given rule was changed.
Moderator conflicts, platform-wide policy shifts, or real-world events can all cause rule changes for reasons outside a single community's control. Future work could attempt to differentiate between rules changed as a result of internal \textit{vs.} external factors.
}

We measure perceptions of governance using public discussion threads, however public discussion threads do not necessarily align with privately held attitudes about governance. Furthermore, community attitudes are not and should not be the only objective when considering rule changes. We also measure discussions about governance and moderators generally, not rules specifically. Future work could develop a more granular classification system.

\new{
Finally, while we use causal inference methods to adjust for two key confounding factors (\sect\ref{sec:iptw}), it is highly likely that there are additional confounding factors that we do not adjust for. Future research could control for  some of these confounding factors, such as content removal rates and moderator workload. Furthermore, future research could use even more robust methods such as randomized controlled trials to avoid bias due to unobserved confounding. Additional data collection could enable time-series based difference-in-difference analyses of rule additions.
}

\subsection{Conclusion}\label{sec:conclusion}
Rules and their enforcement are a critical component of the governance of online communities, yet it is difficult for community moderators to know which rules to select for their community. In this work, we conducted the largest-to-date analysis of rules on Reddit, examining 67,545 unique rules across 5,225 communities over a 5+ year period. We reconstructed timelines of how rules change over time (\sect\ref{sec:timelines}) and classify rules according to their tone, target, and topic (\sect\ref{sec:codebook}). We assessed what types of rules are most prevalent, and how these rules vary across communities of different types (\sect\ref{sec:rq1}). Ours is the first study to connect rules to community outcomes, using data about how community members perceive the governance of their communities. Using these data, we identified the rules most strongly associated with positive community perceptions of governance: Prescriptive Rules (`do this'), User-related rules, and rules about Tags \& Flairs and Peer Engagement (\sect\ref{sec:rq2}).
We also conducted a longitudinal study of the impact of adding new rules to communities, finding that after a rule is added, community perceptions of governance immediately improve, yet this effect `wears off' after approximately six months (\sect\ref{sec:rq3}). Our results have important implications for moderators and community leaders (\sect\ref{sec:discussion}), and we make our datasets public to support future research on this topic.
\section*{Acknowledgments}
This research was supported by the Office of Naval Research (\#N00014-21-1-2154), NSF grant IIS-1901386, NSF CAREER IIS-2142794, NSF grant CNS-2025022, and the Bill \& Melinda Gates Foundation (INV-004841). This work was completed on Hyak, UW’s computing cluster. 

{\small \bibliography{bibliography}}

\begin{thebibliography}{50}
\providecommand{\natexlab}[1]{#1}

\bibitem[{Almerekhi, Kwak, and Jansen(2020)}]{almerekhi_2020_mod_harassment_modeling}
Almerekhi, H.; Kwak, H.; and Jansen, B.~J. 2020.
\newblock Statistical Modeling of Harassment against Reddit Moderators.
\newblock \emph{WWW Companion}.

\bibitem[{Althoff et~al.(2016)Althoff, Sosic, Hicks, King, Delp, and Leskovec}]{althoff2016quantifying}
Althoff, T.; Sosic, R.; Hicks, J.; King, A.; Delp, S.; and Leskovec, J. 2016.
\newblock Quantifying dose response relationships between physical activity and health using propensity scores.
\newblock \emph{NIPS ML4H}.

\bibitem[{Austin and Stuart(2015)}]{Austin2015_IPTW_best_practice}
Austin, P.~C.; and Stuart, E.~A. 2015.
\newblock Moving towards best practice when using inverse probability of treatment weighting (IPTW) using the propensity score to estimate causal treatment effects in observational studies.
\newblock \emph{Statistics in Medicine}, 34.

\bibitem[{Bao et~al.(2021)Bao, Wu, Zhang, Chandrasekharan, and Jurgens}]{Bao_2021_prosocial}
Bao, J.; Wu, J.; Zhang, Y.; Chandrasekharan, E.; and Jurgens, D. 2021.
\newblock Conversations Gone Alright: Quantifying and Predicting Prosocial Outcomes in Online Conversations.
\newblock \emph{TheWebConf}.

\bibitem[{Baumgartner et~al.(2020)Baumgartner, Zannettou, Keegan, Squire, and Blackburn}]{baumgartner_pushshift_2020}
Baumgartner, J.; Zannettou, S.; Keegan, B.; Squire, M.; and Blackburn, J. 2020.
\newblock The Pushshift Reddit Dataset.
\newblock arXiv:2001.08435.

\bibitem[{Chandrasekharan et~al.(2018)Chandrasekharan, Samory, Jhaver, Charvat, Bruckman, Lampe, Eisenstein, and Gilbert}]{eshwar_hiddenrules_2018}
Chandrasekharan, E.; Samory, M.; Jhaver, S.; Charvat, H.; Bruckman, A.; Lampe, C.; Eisenstein, J.; and Gilbert, E. 2018.
\newblock The Internet's Hidden Rules.
\newblock \emph{CSCW}.

\bibitem[{Cima et~al.(2024)Cima, Trujillo, Avvenuti, and Cresci}]{Cima2024TheGB}
Cima, L.; Trujillo, A.; Avvenuti, M.; and Cresci, S. 2024.
\newblock The Great Ban: Efficacy and Unintended Consequences of a Massive Deplatforming Operation on Reddit.
\newblock \emph{Companion Publication of the 16th ACM Web Science Conference}.

\bibitem[{Datta and Adar(2018)}]{Datta2018ExtractingIC}
Datta, S.; and Adar, E. 2018.
\newblock Extracting Inter-community Conflicts in Reddit.
\newblock \emph{ICWSM}.

\bibitem[{Dettmers et~al.(2023)Dettmers, Pagnoni, Holtzman, and Zettlemoyer}]{dettmers_QLoRA_2023}
Dettmers, T.; Pagnoni, A.; Holtzman, A.; and Zettlemoyer, L. 2023.
\newblock QLoRA: Efficient Finetuning of Quantized LLMs.

\bibitem[{Fang, Yang, and Zhu(2023)}]{fang_2023_shaping}
Fang, A.; Yang, W.; and Zhu, H. 2023.
\newblock Shaping Online Dialogue: Examining How Community Rules Affect Discussion Structures on Reddit.
\newblock arXiv:2308.01257.

\bibitem[{Fiesler et~al.(2018)Fiesler, Jiang, McCann, Frye, and Brubaker}]{fiesler_redditrules_2018}
Fiesler, C.; Jiang, J.~A.; McCann, J.; Frye, K.; and Brubaker, J.~R. 2018.
\newblock Reddit Rules! Characterizing an Ecosystem of Governance.
\newblock \emph{ICWSM}.

\bibitem[{Frey and Sumner(2018)}]{Frey2018EmergenceOI}
Frey, S.; and Sumner, R.~W. 2018.
\newblock Emergence of integrated institutions in a large population of self-governing communities.
\newblock \emph{PLoS ONE}, 14.

\bibitem[{Frey et~al.(2022)Frey, Zhong, Bulat, Weisman, Liu, Fujimoto, Wang, and Schweik}]{Frey2022GoverningOG}
Frey, S.; Zhong, Q.; Bulat, B.; Weisman, W.~D.; Liu, C.; Fujimoto, S.; Wang, H.~M.; and Schweik, C.~M. 2022.
\newblock Governing Online Goods: Maturity and Formalization in Minecraft, Reddit, and World of Warcraft Communities.
\newblock \emph{CSCW}.

\bibitem[{Grimmelmann(2015)}]{grimmelmann_2015_virtues_of_mod}
Grimmelmann, J. 2015.
\newblock The Virtues of Moderation.
\newblock \emph{Yale Journal of Law and Technology}.

\bibitem[{Hwang and Foote(2021)}]{hwang_2021_small_communities}
Hwang, S.; and Foote, J.~D. 2021.
\newblock Why Do People Participate in Small Online Communities?
\newblock \emph{CSCW}.

\bibitem[{Hwang and Shaw(2022)}]{Hwang2022RulesAR}
Hwang, S.; and Shaw, A. 2022.
\newblock Rules and Rule-Making in the Five Largest Wikipedias.
\newblock \emph{ICWSM}.

\bibitem[{Jhaver et~al.(2019)Jhaver, Appling, Gilbert, and Bruckman}]{jhaver_suspectremoved_2019}
Jhaver, S.; Appling, D.~S.; Gilbert, E.; and Bruckman, A. 2019.
\newblock "Did you suspect the post would be removed?" Understanding user reactions to content removals on Reddit.
\newblock \emph{CSCW}.

\bibitem[{Jhaver, Bruckman, and Gilbert(2019)}]{jhaver_2019_removal_reasons}
Jhaver, S.; Bruckman, A.; and Gilbert, E. 2019.
\newblock Does Transparency in Moderation Really Matter?
\newblock \emph{CSCW}.

\bibitem[{Kairam, Mercado, and Sumner(2022)}]{Kairam_2022_twitch_sovc}
Kairam, S.~R.; Mercado, M.~C.; and Sumner, S.~A. 2022.
\newblock A Social-Ecological Approach to Modeling Sense of Virtual Community (SOVC) in Livestreaming Communities.
\newblock \emph{CSCW}.

\bibitem[{Keegan and Fiesler(2017)}]{keegan_wikirules_2017}
Keegan, B.; and Fiesler, C. 2017.
\newblock The Evolution and Consequences of Peer Producing Wikipedia’s Rules.
\newblock \emph{ICWSM}.

\bibitem[{Kittur and Kraut(2010)}]{kittur_2010_wikia}
Kittur, A.; and Kraut, R.~E. 2010.
\newblock Beyond Wikipedia: coordination and conflict in online production groups.
\newblock \emph{CSCW}.

\bibitem[{Koshy et~al.(2023)Koshy, Bajpai, Chandrasekharan, Sundaram, and Karahalios}]{koshy_2023_user_mod_alignment}
Koshy, V.; Bajpai, T.; Chandrasekharan, E.; Sundaram, H.; and Karahalios, K. 2023.
\newblock Measuring User-Moderator Alignment on r/ChangeMyView.
\newblock \emph{CSCW}.

\bibitem[{Kraut and Resnick(2012)}]{kraut_communities_2012}
Kraut, R.~E.; and Resnick, P. 2012.
\newblock \emph{Building Successful Online Communities}.
\newblock MIT Press.

\bibitem[{Landis and Koch(1977)}]{landis_cohenkappa_1977}
Landis, J.~R.; and Koch, G.~G. 1977.
\newblock The Measurement of Observer Agreement for Categorical Data.
\newblock \emph{Biometrics}.

\bibitem[{Li, Hecht, and Chancellor(2022)}]{li_2022_modlogs}
Li, H.; Hecht, B.~J.; and Chancellor, S. 2022.
\newblock All That's Happening behind the Scenes: Putting the Spotlight on Volunteer Moderator Labor in Reddit.
\newblock \emph{ICWSM}.

\bibitem[{Liu et~al.(2019)Liu, Ott, Goyal, Du, Joshi, Chen, Levy, Lewis, Zettlemoyer, and Stoyanov}]{liu_roberta_2019}
Liu, Y.; Ott, M.; Goyal, N.; Du, J.; Joshi, M.; Chen, D.; Levy, O.; Lewis, M.; Zettlemoyer, L.; and Stoyanov, V. 2019.
\newblock RoBERTa: {A} Robustly Optimized {BERT} Pretraining Approach.
\newblock arXiv:1907.11692.

\bibitem[{{Llama Team}(2024)}]{dubey_llama3herd_2024}
{Llama Team}. 2024.
\newblock The Llama 3 Herd of Models.
\newblock arXiv:2407.21783.

\bibitem[{Lloyd et~al.(2024)Lloyd, Gosciak, Nguyen, and Naaman}]{lloyd_2024_airules}
Lloyd, T.; Gosciak, J.; Nguyen, T.; and Naaman, M. 2024.
\newblock AI Rules? Characterizing Reddit Community Policies Towards AI-Generated Content.
\newblock arXiv:2410.11698.

\bibitem[{MacQueen et~al.(1998)MacQueen, McLellan, Kay, and Milstein}]{MacQueen1998CodebookDF}
MacQueen, K.; McLellan, E.; Kay, K.; and Milstein, B. 1998.
\newblock Codebook Development for Team-Based Qualitative Analysis.
\newblock \emph{Field Methods}, 10: 31 -- 36.

\bibitem[{Matias(2019{\natexlab{a}})}]{matias_2019_civic_labor}
Matias, J.~N. 2019{\natexlab{a}}.
\newblock The Civic Labor of Volunteer Moderators Online.
\newblock \emph{Social Media + Society}.

\bibitem[{Matias(2019{\natexlab{b}})}]{Matias_2019_Preventing}
Matias, J.~N. 2019{\natexlab{b}}.
\newblock Preventing harassment and increasing group participation through social norms in 2,190 online science discussions.
\newblock \emph{PNAS}.

\bibitem[{{MistralAI}(2024)}]{mistralai_mistral-7b-instruct-v03_2024}
{MistralAI}. 2024.
\newblock Mistral-{7B}-{Instruct}-v0.3.

\bibitem[{Nicholson, Keegan, and Fiesler(2023)}]{Nicholson2023MastodonRC}
Nicholson, M.~N.; Keegan, B.~C.; and Fiesler, C. 2023.
\newblock Mastodon Rules: Characterizing Formal Rules on Popular Mastodon Instances.
\newblock \emph{CSCW}.

\bibitem[{Nussbaum et~al.(2024)Nussbaum, Morris, Duderstadt, and Mulyar}]{nussbaum_nomicembed_2024}
Nussbaum, Z.; Morris, J.~X.; Duderstadt, B.; and Mulyar, A. 2024.
\newblock Nomic Embed: Training a Reproducible Long Context Text Embedder.
\newblock arXiv:2402.01613.

\bibitem[{OpenAI(2024)}]{openai_gpt4technicalreport_2024}
OpenAI. 2024.
\newblock GPT-4 Technical Report.
\newblock arXiv:2303.08774.

\bibitem[{{OpenAI}(2024)}]{openai_hello_2024}
{OpenAI}. 2024.
\newblock Hello {GPT}-4o.

\bibitem[{Prinster et~al.(2024)Prinster, Smith, Tan, and Keegan}]{prinster_2024_archetypes}
Prinster, G.~H.; Smith, C.~E.; Tan, C.; and Keegan, B.~C. 2024.
\newblock Community Archetypes: An Empirical Framework for Guiding Research Methodologies to Reflect User Experiences of Sense of Virtual Community on Reddit.
\newblock \emph{CSCW}.

\bibitem[{Proferes et~al.(2021)Proferes, Jones, Gilbert, Fiesler, and Zimmer}]{Proferes_2021_reddit_research_overview}
Proferes, N.; Jones, N.; Gilbert, S.~A.; Fiesler, C.; and Zimmer, M. 2021.
\newblock Studying Reddit: A Systematic Overview of Disciplines, Approaches, Methods, and Ethics.
\newblock \emph{Social Media + Society}.

\bibitem[{Reddy and Chandrasekharan(2023)}]{Reddy2023EvolutionOR}
Reddy, H.; and Chandrasekharan, E. 2023.
\newblock Evolution of Rules in Reddit Communities.
\newblock \emph{CSCW}.

\bibitem[{Ribeiro et~al.(2020)Ribeiro, Jhaver, Zannettou, Blackburn, Cristofaro, Stringhini, and West}]{Ribeiro2020_migration}
Ribeiro, M.~H.; Jhaver, S.; Zannettou, S.; Blackburn, J.; Cristofaro, E.~D.; Stringhini, G.; and West, R. 2020.
\newblock Do Platform Migrations Compromise Content Moderation? Evidence from r/The\_Donald and r/Incels.
\newblock \emph{CSCW}.

\bibitem[{Seering and Kairam(2022)}]{seering_2022_twitch_moderator_recruiting}
Seering, J.; and Kairam, S.~R. 2022.
\newblock Who Moderates on Twitch and What Do They Do?
\newblock \emph{GROUP}.

\bibitem[{Srinivasan et~al.(2019)Srinivasan, Danescu-Niculescu-Mizil, Lee, and Tan}]{srinivasan_2019_content_removal_cmv}
Srinivasan, K.~B.; Danescu-Niculescu-Mizil, C.; Lee, L.; and Tan, C. 2019.
\newblock Content Removal as a Moderation Strategy.
\newblock \emph{CSCW}.

\bibitem[{TeBlunthuis et~al.(2022)TeBlunthuis, Kiene, Brown, Levi, McGinnis, and Hill}]{TeBlunthuis_2022_similar_communities}
TeBlunthuis, N.; Kiene, C.; Brown, I.; Levi, L.; McGinnis, N.; and Hill, B.~M. 2022.
\newblock No Community Can Do Everything: Why People Participate in Similar Online Communities.
\newblock \emph{CSCW}.

\bibitem[{{The Internet Archive}(2025)}]{wayback_machine}
{The Internet Archive}. 2025.
\newblock Https://web.archive.org/.

\bibitem[{Thomas et~al.(2021)Thomas, Riehm, Glenski, and Weninger}]{Thomas_2021_bans_behavior}
Thomas, P.~B.; Riehm, D.; Glenski, M.; and Weninger, T. 2021.
\newblock Behavior Change in Response to Subreddit Bans and External Events.
\newblock \emph{IEEE TCSS}.

\bibitem[{Weld et~al.(2024)Weld, Leibmann, Zhang, and Althoff}]{weld_perceptions_2024}
Weld, G.; Leibmann, L.; Zhang, A.~X.; and Althoff, T. 2024.
\newblock Perceptions of Moderators as a Large-Scale Measure of Online Community Governance.

\bibitem[{Weld, Zhang, and Althoff(2021)}]{weld_surveys_2021}
Weld, G.; Zhang, A.~X.; and Althoff, T. 2021.
\newblock What Makes Online Communities 'Better'? Measuring Values, Consensus, and Conflict across Thousands of Subreddits.

\bibitem[{Weld, Zhang, and Althoff(2024)}]{Weld_2024_taxonomy}
Weld, G.; Zhang, A.~X.; and Althoff, T. 2024.
\newblock Making Online Communities 'Better': A Taxonomy of Community Values on Reddit.
\newblock \emph{ICWSM}.

\bibitem[{Zhang, Hugh, and Bernstein(2020)}]{Zhang2020PolicyKitBG}
Zhang, A.~X.; Hugh, G.; and Bernstein, M.~S. 2020.
\newblock PolicyKit: Building Governance in Online Communities.
\newblock \emph{UIST}.

\bibitem[{Zhu, Kraut, and Kittur(2014)}]{zhu_2014_wikia_membership}
Zhu, H.; Kraut, R.~E.; and Kittur, A. 2014.
\newblock The Impact of Membership Overlap on the Survival of Online Communities.
\newblock \emph{CHI}.

\end{thebibliography}

\clearpage
\section*{Ethics Checklist}


\begin{enumerate}
    \item  Would answering this research question advance science without violating social contracts, such as violating privacy norms, perpetuating unfair profiling, exacerbating the socio-economic divide, or implying disrespect to societies or cultures?
    \answerYes{Yes, we believe that our work makes important contributions to the field of computational social science with clear implications and broader impact (\sect\ref{sec:discussion}), and we take steps to address bias and protect privacy of our participants (\sect\ref{sec:ethics}).}
  \item Do your main claims in the abstract and introduction accurately reflect the paper's contributions and scope?
    \answerYes{Yes, all claims made are concretely supported.}
   \item Do you clarify how the proposed methodological approach is appropriate for the claims made? 
    \answerYes{Yes, our methodology is clear and rational given claims made}
   \item Do you clarify what are possible artifacts in the data used, given population-specific distributions?
    \answerYes{Yes, All artifacts are stated and addressed}
  \item Did you describe the limitations of your work?
    \answerYes{Yes, Limitations are discussed in \ref{sec:limits}}
  \item Did you discuss any potential negative societal impacts of your work?
    \answerYes{Yes, negative impacts are discussed in \sect\ref{sec:ethics}.}
  \item Did you discuss any potential misuse of your work?
    \answerYes{Yes, misuse is discussed in \sect\ref{sec:ethics}.}
  \item Did you describe steps taken to prevent or mitigate potential negative outcomes of the research, such as data and model documentation, data anonymization, responsible release, access control, and the reproducibility of findings?
    \answerYes{Yes, data is aggregated and anonymized to remove sources of harm. This is described in detail in Appendix~\ref{app:dataset_ethics}}
  \item Have you read the ethics review guidelines and ensured that your paper conforms to them?
    \answerYes{Yes, this paper conforms to ethics review guidelines.}
  \item Did you clearly state the assumptions underlying all theoretical results?
    \answerNA{Not applicable}
  \item Have you provided justifications for all theoretical results?
    \answerNA{Not applicable}
  \item Did you discuss competing hypotheses or theories that might challenge or complement your theoretical results?
    \answerNA{Not applicable}
  \item Have you considered alternative mechanisms or explanations that might account for the same outcomes observed in your study?
    \answerYes{Yes, alternative mechanisms are discussed in \sect\ref{sec:limits}.}
  \item Did you address potential biases or limitations in your theoretical framework?
    \answerYes{Yes, bias and limitations are discussed in \sect\ref{sec:limits}.}
  \item Have you related your theoretical results to the existing literature in social science?
    \answerNA{Not applicable}
  \item Did you discuss the implications of your theoretical results for policy, practice, or further research in the social science domain?
    \answerYes{Yes, we discuss implications and future directions in \sect\ref{sec:discussion}}

  \item Did you state the full set of assumptions of all theoretical results?
    \answerNA{Not applicable}
   \item Did you include complete proofs of all theoretical results?
    \answerNA{Not applicable}

  \item Did you include the code, data, and instructions needed to reproduce the main experimental results (either in the supplemental material or as a URL)?
    \answerYes{Code and data are available online at \texttt{\tiny bdata.cs.washington.edu/mod-perceptions}}
  \item Did you specify all the training details (e.g., data splits, hyperparameters, how they were chosen)?
    \answerYes{Yes, all training and prompting details are specified in Appendix~\ref{app:prompt}.}
     \item Did you report error bars (e.g., with respect to the random seed after running experiments multiple times)?
    \answerYes{Yes, bootstrapped 95\% confidence intervals are reported for all results.}
	\item Did you include the total amount of compute and the type of resources used (e.g., type of GPUs, internal cluster, or cloud provider)?
    \answerYes{Yes, compute details are given in \sect\ref{sec:data}.}
     \item Do you justify how the proposed evaluation is sufficient and appropriate to the claims made? 
    \answerYes{Yes, discussed in \sect\ref{sec:data}}
     \item Do you discuss what is ``the cost`` of misclassification and fault (in)tolerance?
    \answerYes{Yes, the cost of misclassification is addressed in \sect\ref{sec:limits}}
  \item If your work uses existing assets, did you cite the creators?
    \answerYes{Yes, our work makes extensive use of existing models, among other assets. All are cited}
  \item Did you mention the license of the assets?
    \answerYes{Assets were shared without formal licensing beyond a request to cite relevant papers, which we comply with.}
  \item Did you include any new assets in the supplemental material or as a URL?
    \answerYes{Code and data are available online at \texttt{\tiny bdata.cs.washington.edu/mod-perceptions}}
  \item Did you discuss whether and how consent was obtained from people whose data you're using/curating?
    \answerYes{Yes, source data licenses are discussed}
  \item Did you discuss whether the data you are using/curating contains personally identifiable information or offensive content?
    \answerYes{Yes, the risks of deanonymization and improper conduct are discussed}
    \item If you are curating or releasing new datasets, did you discuss how you intend to make your datasets FAIR?
        \answerYes{Yes, FAIR guidelines are present in Appendix~\ref{app:dataset_ethics}}
    \item If you are curating or releasing new datasets, did you create a Datasheet for the Dataset? 
        \answerYes{Yes, Datasheet is included in Appendix~\ref{app:dataset_ethics}}
\end{enumerate}

\clearpage
\onecolumn
\appendix
\begin{center}
     \huge \textbf{Appendix}
\end{center}

\section{Codebook}\label{app:codebook_long}

We provide a copy of our codebook, with further details and examples, as shared among authors for study reproduction. This codebook was created iteratively until saturation, with further detail found in \sect\ref{sec:codebook}.

\subsection{Rule Tone}

\xhdr{Prescriptive}
A rule expressing general guidelines or desires for a community.
\begin{itemize}
    \item Ex. 6. Include [REQUEST] in text posts for gif requests.
    \item Ex.  Remember your reddiquette.
\end{itemize}

\xhdr{Restrictive}
A rule which explicitly limits or forbids certain actions.
\begin{itemize}
    \item Ex. Don’t be an ass.
    \item Ex.  Low effort \& low quality content.
\end{itemize}

\subsection{Rule Target}
Rules generally target one or more of the following.

\xhdr{Post Content}
Any rule explicitly stating desired or undesired content within the subreddit.
\begin{itemize}
    \item Ex. 2. Text body must include context.
    \item Ex. OC is welcome.
    \item Ex. 1. No advertising, marketing research and spam.
    \item Ex. 5. Low effort/Blog spam/Repost - mods, please review.
\end{itemize}

\xhdr{User-Related}
Any rule related to users, or which would cause unequal enforcement if two different users posted the same content. This includes verification and prior approval rules.
\begin{itemize}
    \item Ex. Submissions from new accounts or from accounts with fewer than 10 karma will be removed.
    \item Ex. IamAs/AMAs must be approved by mods.
\end{itemize}

\xhdr{Post Format}
Any rule prescribing post structure, formatting, titling, tagging or referencing a location to post (other subreddits, specific threads).
\begin{itemize}
    \item Ex. 3. All News And Tweets Should Follow Standard Format.
    \item Ex. 2. R2: BootTooBig format.
    \item Ex. 10. Posts must begin with "ELI5:"
    \item Ex. 8. Use the stickied Megathreads If there are stickied Megathreads provided about certain topics they must be used when discussing that topic. Other posts will be removed.
\end{itemize}

\xhdr{Not a Rule}
Sidebar content that is not necessarily a rule.

\begin{itemize}
    \item Ex. Welcome to r/guitar, a community devoted to the exchange of guitar related information...
    \item Ex. Discuss all the Real Housewives franchises by Bravo TV with us...
    \item Ex. 9. Have fun!
\end{itemize}

\subsection{Rule Topic}
A rule can have any number of specific topics of focus.

\xhdr{Post Tagging \& Flairing}
Any rule pertaining to labeling/flairing, marking nsfw, using tags, etc.
\begin{itemize}
    \item Ex. If your post is NSFW, please label it as such.
    \item Ex. Please add flair to your submissions.
    \item Ex. Use spoiler tags appropriately.
\end{itemize}

\xhdr{Peer Engagement}
Any rule encouraging or discouraging the quantity of on-site peer engagement, including reporting, commenting, voting, and general activity.

\begin{itemize}
    \item Ex. Please report any rule-breaking posts, as well as abusive comments or harassment.
    \item Ex. The downvote button is NOT a disagree button. 
    \item Ex. Upvote Begging.
    \item Ex. If posting about weight loss, please provide details of your current plan and respond to questions.
\end{itemize}

\xhdr{Brigading}
Any rule regulating large group actions, including raiding or mass-voting.

\begin{itemize}
    \item Ex. This subreddit will not participate in or be the source of brigades or raids on other subreddits.
    \item Ex. Vote brigading.
    \item Ex. Use NP Links When Posting A Link To Other Subs.
\end{itemize}

\xhdr{Links \& External Content}
Any rule regulating links or the content of external resources. These rules do not regulate the content of images, but can regulate the host thereof.

\begin{itemize}
    \item Ex. Do not submit a shortened link using a URL shortener like tinyurl.
    \item Ex. No promoting other subs/discords/crews.
    \item Ex. Do not mention or ask for usernames for other services.
\end{itemize}

\xhdr{Images}
Any rule pertaining to the content or quality of images or videos, often including memes.

\begin{itemize}
    \item Ex. While the places posted should be aesthetically unappealing, it is recommended that the photo quality is good. Artsy shots are more than welcome.
    \item Ex. Picture quality.
    \item Ex. Sourcing fan-art.
    \item Ex. Rule 4 - No memes, macros, low-effort, rant posts.
\end{itemize}

\xhdr{Commercialization}
Any rule regulating advertisement, self-promotion, referrals and referral links, or buying/selling.

\begin{itemize}
    \item Ex. Limit self-promotion.
    \item Ex. Do not advertise products or groups without permission.
    \item Ex. No VPN or Crypto-Currency Discussions Due to the commercial nature of VPNs...
    \item Ex. Rule 18 - No product/store/page/app review/rating.
\end{itemize}

\xhdr{Illegal Content}
Rules about illegal content, including copyright infringement and piracy. Does not include harassment and hate speech. 

\begin{itemize}
    \item Ex. No talking about where to buy or get hold of MDMA or other illegal drugs.
    \item Ex. No links to pirated materials.
    \item Ex. No discussion of theft whatsoever.
\end{itemize}

\xhdr{Divisive Content}
Rules regulating content which is inherently divisive, such as politics, current events, or community-specific hot topics. This only includes specific topics, i.e, the what, not the how.

\begin{itemize}
    \item Ex. Ukraine Conflict.
    \item Ex. Religious preaching.
    \item Ex. Politics / Current Events.
    \item Ex. Do Not Editorialize.
\end{itemize}

\xhdr{Respect for Others} 
Rules explicitly discouraging hate speech, antagonization, harassment or encouraging respect for others. Also covers rules about swearing, which is often disrespectful.
\begin{itemize}
    \item Ex. Any comments that include hate speech will be deleted.
    \item Ex. No hate speech.
    \item Ex. No profane/vulgar/undignified language.
    \item Ex. Social jerking.
\end{itemize}

\xhdr{Spam, Low Quality, Off-Topic, and Reposts}
Any rule covering low-quality, off-topic, spam, reposted, or otherwise generally-undesired content.

\begin{itemize}
    \item Ex. No advertising, marketing research and spam.
    \item Ex. Low effort/Blog spam/Repost - mods, please review!
    \item Ex. Common reposts and overdone topics will be removed.
    \item Ex. Embargoed Topics.
    \item Ex. Your post must be an unpopular opinion.
\end{itemize}

\xhdr{Ban Mentioned}
Any rule that explicitly mentions banning.

\begin{itemize}
    \item Ex. Malicious attempts to spoil other users will result in a ban. 
\end{itemize}

\xhdr{Karma/Score/Voting Mentioned}
Any rule that explicitly mentions karma, user scores, or voting.

\begin{itemize}
    \item Ex. No Pandering for Upvotes.
\end{itemize}

\clearpage
\section{Classification Prompt}\label{app:prompt}

The prompt was created collaboratively and iterated on to maximize classification performance across a manually-labeled validation set of 100 examples. Few-shot examples were chosen from a train-set of 200 randomly-sampled rules. All large-language models evaluated use the same prompt and sample selection technique. See \sect\ref{sec:codebook} for details.

\begin{tcolorbox}[enhanced,
  colframe=mustard,
  colback=gray!3,
  frame style={draw=black, line width=0.2mm},
  rounded corners,
  width=\textwidth,
  breakable
]
\small
{\ttfamily
Given a rule in a specific subreddit, identify topics and qualities about the rule.
\\\\*
If the rule explicitly limits or forbids certain actions, mark it as "Restrictive".
\\*
If a rule expresses general guidelines or desires for a community, mark it as "Prescriptive".
\\*
If a rule explicitly states desired or undesired content within the subreddit, mark it as "Post Content".
\\*
If a rule is related to users or would cause unequal enforcement if two different users posted the same content (including verification and prior approval rules), mark it as "User-Related".
\\*
If a rule prescribes post structure, formatting, titling, or references a location to post (such as other subreddits or specific threads), mark it as "Post Format".
\\*
If the content is sidebar information and not necessarily a rule, mark it as "Not a Rule".
\\*
If a rule pertains to labeling/flairing, marking nsfw, using tags, etc., mark it as "Post Tagging \& Flairing".
\\*
If a rule encourages or discourages the quantity of on-site peer engagement, including reporting, commenting, voting, and general activity, mark it as "Peer Engagement".
\\*
If a rule regulates large group actions, including raiding or mass-voting, mark it as "Brigading".
\\*
If a rule is about links or the external, off-reddit content (excluding image content), mark it as "Links \& External Content".
\\*
If a rule pertains to the content or quality of images or videos, often including memes, mark it as "Images".
\\*
If a rule regulates advertisement, self-promotion, referrals and referral links, or buying/selling, mark it as "Commercialization".
\\*
If a rule explicitly mentions illegal content, including copyright infringement and piracy (but excluding harassment and hate speech), mark it as "Illegal Content".
\\*
If a rule regulates content that is inherently divisive, such as politics, current events, or community-specific hot topics. , mark it as "Divisive Content". Only mark if the rule covers specific topics, i.e, the what being said, not how it was said.
\\*
If a rule discourages hate speech, antagonization, harassment, or encourages respect for others (including swearing), mark it as "Respect for Others".
\\*
If a rule covers low-quality, off-topic, spam, reposted, or otherwise generally-undesired content, mark it as "Spam, Low Quality, Off-Topic, and Reposts".
\\*
If a rule explicitly mentions banning users, mark it as "Ban Mentioned".
\\*
If a rule explicitly mentions karma or user scores, mark it as "Karma/Score Mentioned".
}
\\*
{\ttfamily 
Your answer should follow the format given in the examples:
\begin{verbatim}
subreddit : AskReddit
rule : 3. Rule 3 - Open ended questions only
rule_description : 
Prescriptive : False
Restrictive : True
Post Content : False
Post Format : True
User-Related : False
Not a Rule : False
Spam, Low Quality, Off-Topic, and Reposts : False
Post Tagging & Flairing : False
Peer Engagement : False
Links & External Content : False
Images : False
Commercialization : False
Illegal Content : False
Divisive Content : False
Respect for Others : False
Brigading : False
Ban Mentioned : False
Karma/Score Mentioned : False
\end{verbatim}
}
{... 5 more examples formatted as above}
\end{tcolorbox}

\clearpage
\section{Additional Details on IPTW, Covariates, \& Balance}\label{app:iptw_balance}

\new{
As described in \sect\ref{sec:iptw}, we use Inverse Probability of Treatment Weighing (IPTW) \cite{Austin2015_IPTW_best_practice} to adjust for two key confounding factors, community topic and size, in our analyses in \sect\ref{sec:rq2}. We use IPTW to reweight observations in the treatment (having a rule of a given type) and control (lacking a rule of a given type) groups to make their weighted distributions of covariates more similar to the reference distribution, consisting of the entire population \cite{althoff2016quantifying}. For more details, see \sect\ref{sec:iptw}.
}

\new{
To understand the degree to which IPTW changes our experimental results from those computed without any confounding adjustment (non-adjusted/unweighted), the following figure shows both IPTW results (in blue and red) and non-adjusted results (in gray).
In these analyses, IPTW mostly slightly decreases the estimated effect size while not zeroing it out. This suggests that while some observed differences between communities' perceptions of governance can be partially, but only partially, be explained by community size and topic, and therefore they suggest that rules also play an important role.
}

\begin{figure*}[hbp]
   \centering
   \includegraphics[width=1\textwidth]{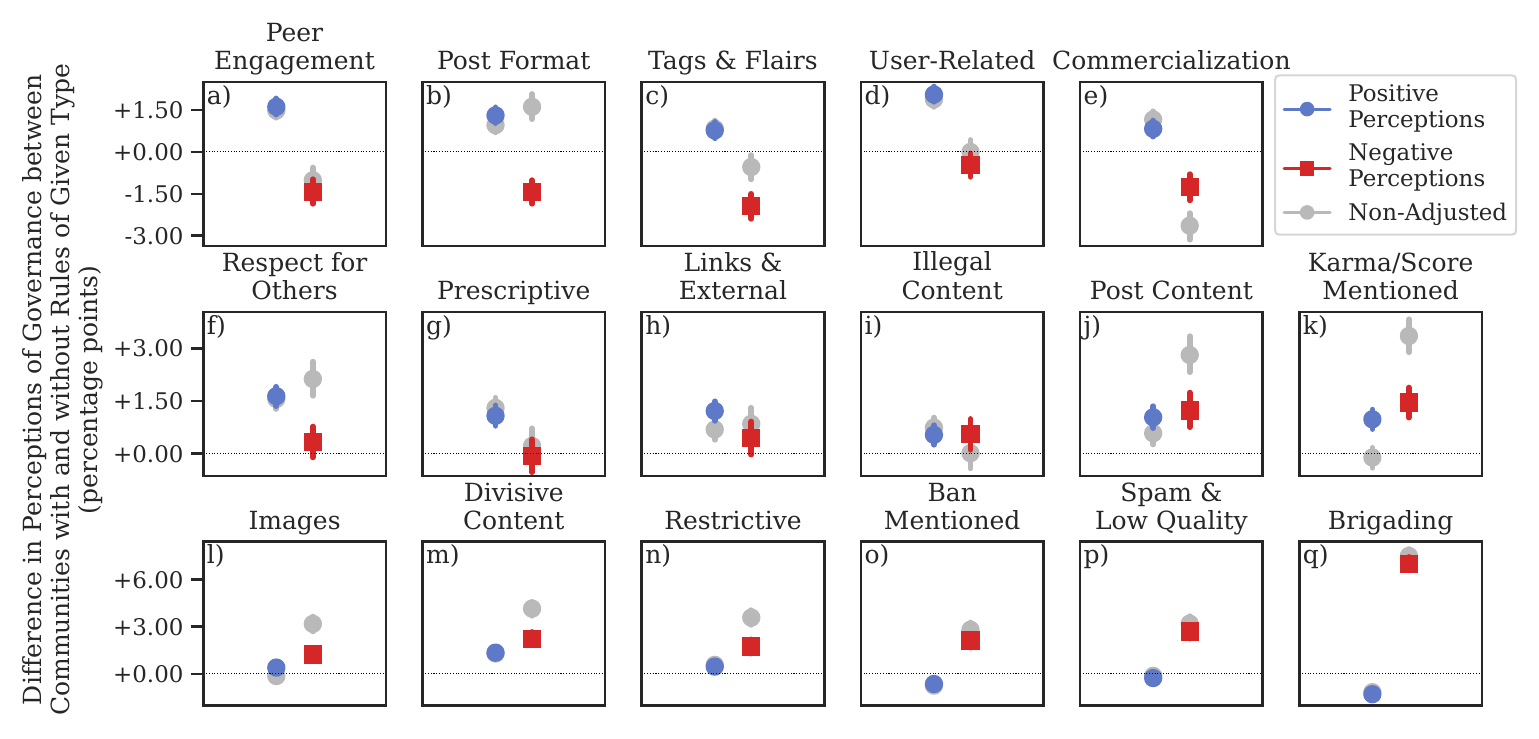}
   \vspace{-8mm}
 \caption{\new{For most rule types, adjusting for community topic and size (blue and red markers) slightly reduces the difference between communities with and without different rules, compared to no adjustment for confounding (gray markers). This suggests that topic size partially, but only partially, explain some difference in perceptions of governance, and that rules play an important rule. Additional discussion of these results is provided in \sect\ref{sec:rq2}.}
}
 \label{fig:iptw_with_unweighted}
\end{figure*}

\new{
For IPTW computation, we measure Community Size by computing the average number of posts and comments uploaded to a given community each day during the study period (April 2018 to December 2023). Community Topic is classified into one of six different topical classes provided by an existing dataset \cite{Weld_2024_taxonomy} and are one-hot encoded for our logistic regression propensity score model. As such, the covariate vector $\mathbf{X}_i$ is of length 7. 
}


\new{
We assess the efficacy of IPTW by examining the standard mean difference (SMD) between each weighted treatment/control group and the reference distribution, consisting of the entire population. Two groups are often considered `balanced' or `indistinguishable` if all covariates are within an SMD of 0.25 from the reference distribution \cite{althoff2016quantifying}.
Across 238 condition-covariate-rule type pairs (17 experiments $\times$ 2 treatment/control conditions $\times$ 7 covariates), our method achieves balance in 235 cases (98.74\%). 
In the three cases where our method fails to achieve balance, no experiment has more than a single unbalanced covariate (out of seven), and no SMD exceeds 1.00. A complete list of covariates and SMDs for each experiment is given the in the tables below, with SMDs greater than 0.25 indicated with bold text.
}

\subsection{Covariate Balance for Prescriptive Rules}
\begin{table}[H]
\centering
\label{tab:balance_prescriptive}
\begin{tabular}{lccccc}
\toprule
Covariate & \makecell{Mean of Reference\\Population} & \makecell{Mean of Control\\Group (no rule)} & \makecell{Mean of Treated\\Group (has rule)} & SMD for Control & SMD for Treated \\
\midrule
Community Size & 1661.26 & 1289.02 & 1781.88 & \textbf{0.46} & 0.00 \\
Discussion & 0.11 & 0.11 & 0.11 & 0.02 & 0.01 \\
Hobby & 0.51 & 0.48 & 0.52 & -0.02 & 0.00 \\
Identity & 0.17 & 0.17 & 0.17 & 0.00 & 0.00 \\
Media & 0.14 & 0.16 & 0.13 & 0.02 & -0.01 \\
Memes & 0.05 & 0.05 & 0.05 & -0.04 & 0.01 \\
News & 0.02 & 0.03 & 0.02 & 0.03 & -0.01 \\
\bottomrule
\end{tabular}
\end{table}

\subsection{Covariate Balance for Restrictive Rules}
\begin{table}[H]
\centering
\label{tab:balance_restrictive}
\begin{tabular}{lccccc}
\toprule
Covariate & \makecell{Mean of Reference\\Population} & \makecell{Mean of Control\\Group (no rule)} & \makecell{Mean of Treated\\Group (has rule)} & SMD for Control & SMD for Treated \\
\midrule
Community Size & 1661.26 & 379.22 & 1798.68 & -0.04 & 0.00 \\
Discussion & 0.11 & 0.11 & 0.11 & 0.10 & 0.00 \\
Hobby & 0.51 & 0.53 & 0.51 & 0.00 & 0.00 \\
Identity & 0.17 & 0.14 & 0.17 & -0.04 & 0.00 \\
Media & 0.14 & 0.15 & 0.14 & 0.00 & 0.00 \\
Memes & 0.05 & 0.04 & 0.05 & -0.08 & 0.00 \\
News & 0.02 & 0.03 & 0.02 & -0.01 & 0.00 \\
\bottomrule
\end{tabular}
\end{table}

\subsection{Covariate Balance for Post Content Rules}
\begin{table}[H]
\centering
\label{tab:balance_post_content}
\begin{tabular}{lccccc}
\toprule
Covariate & \makecell{Mean of Reference\\Population} & \makecell{Mean of Control\\Group (no rule)} & \makecell{Mean of Treated\\Group (has rule)} & SMD for Control & SMD for Treated \\
\midrule
Community Size & 1661.26 & 258.60 & 1790.65 & -0.12 & 0.00 \\
Discussion & 0.11 & 0.10 & 0.11 & 0.03 & 0.00 \\
Hobby & 0.51 & 0.52 & 0.51 & 0.04 & 0.00 \\
Identity & 0.17 & 0.15 & 0.17 & -0.01 & 0.00 \\
Media & 0.14 & 0.16 & 0.14 & -0.02 & 0.00 \\
Memes & 0.05 & 0.04 & 0.05 & -0.11 & 0.00 \\
News & 0.02 & 0.03 & 0.02 & 0.04 & 0.00 \\
\bottomrule
\end{tabular}
\end{table}

\subsection{Covariate Balance for Post Format Rules}
\begin{table}[H]
\centering
\label{tab:balance_post_format}
\begin{tabular}{lccccc}
\toprule
Covariate & \makecell{Mean of Reference\\Population} & \makecell{Mean of Control\\Group (no rule)} & \makecell{Mean of Treated\\Group (has rule)} & SMD for Control & SMD for Treated \\
\midrule
Community Size & 1661.26 & 707.31 & 2659.29 & 0.14 & 0.01 \\
Discussion & 0.11 & 0.11 & 0.12 & 0.05 & 0.01 \\
Hobby & 0.51 & 0.51 & 0.50 & -0.05 & -0.01 \\
Identity & 0.17 & 0.18 & 0.16 & -0.02 & -0.01 \\
Media & 0.14 & 0.13 & 0.15 & 0.03 & 0.01 \\
Memes & 0.05 & 0.04 & 0.05 & 0.05 & 0.00 \\
News & 0.02 & 0.02 & 0.03 & -0.02 & 0.00 \\
\bottomrule
\end{tabular}
\end{table}

\subsection{Covariate Balance for User-Related Rules}
\begin{table}[H]
\centering
\label{tab:balance_user_related}
\begin{tabular}{lccccc}
\toprule
Covariate & \makecell{Mean of Reference\\Population} & \makecell{Mean of Control\\Group (no rule)} & \makecell{Mean of Treated\\Group (has rule)} & SMD for Control & SMD for Treated \\
\midrule
Community Size & 1661.26 & 1540.82 & 2028.53 & 0.04 & 0.03 \\
Discussion & 0.11 & 0.10 & 0.15 & 0.00 & 0.01 \\
Hobby & 0.51 & 0.52 & 0.48 & -0.01 & 0.01 \\
Identity & 0.17 & 0.16 & 0.19 & 0.00 & -0.01 \\
Media & 0.14 & 0.14 & 0.12 & 0.00 & -0.01 \\
Memes & 0.05 & 0.05 & 0.03 & 0.00 & 0.00 \\
News & 0.02 & 0.02 & 0.02 & 0.00 & -0.01 \\
\bottomrule
\end{tabular}
\end{table}

\subsection{Covariate Balance for Spam, Low Quality, Off-Topic, and Reposts Rules}
\begin{table}[H]
\centering
\label{tab:balance_spam_low_quality_off_topic_and_reposts}
\begin{tabular}{lccccc}
\toprule
Covariate & \makecell{Mean of Reference\\Population} & \makecell{Mean of Control\\Group (no rule)} & \makecell{Mean of Treated\\Group (has rule)} & SMD for Control & SMD for Treated \\
\midrule
Community Size & 1661.26 & 502.99 & 1992.24 & 0.01 & 0.00 \\
Discussion & 0.11 & 0.14 & 0.10 & 0.13 & 0.01 \\
Hobby & 0.51 & 0.51 & 0.51 & -0.05 & -0.01 \\
Identity & 0.17 & 0.17 & 0.17 & 0.01 & 0.00 \\
Media & 0.14 & 0.13 & 0.14 & -0.03 & 0.00 \\
Memes & 0.05 & 0.03 & 0.05 & -0.02 & 0.00 \\
News & 0.02 & 0.02 & 0.03 & -0.04 & 0.01 \\
\bottomrule
\end{tabular}
\end{table}

\subsection{Covariate Balance for Post Tagging \& Flairing Rules}
\begin{table}[H]
\centering
\label{tab:balance_post_tagging_flairing}
\begin{tabular}{lccccc}
\toprule
Covariate & \makecell{Mean of Reference\\Population} & \makecell{Mean of Control\\Group (no rule)} & \makecell{Mean of Treated\\Group (has rule)} & SMD for Control & SMD for Treated \\
\midrule
Community Size & 1661.26 & 1201.43 & 3035.81 & 0.02 & 0.01 \\
Discussion & 0.11 & 0.13 & 0.07 & 0.00 & 0.00 \\
Hobby & 0.51 & 0.50 & 0.55 & 0.00 & 0.00 \\
Identity & 0.17 & 0.19 & 0.10 & 0.00 & 0.00 \\
Media & 0.14 & 0.12 & 0.20 & 0.01 & 0.00 \\
Memes & 0.05 & 0.04 & 0.06 & -0.01 & 0.00 \\
News & 0.02 & 0.03 & 0.02 & 0.03 & -0.01 \\
\bottomrule
\end{tabular}
\end{table}

\subsection{Covariate Balance for Peer Engagement Rules}
\begin{table}[H]
\centering
\label{tab:balance_peer_engagement}
\begin{tabular}{lccccc}
\toprule
Covariate & \makecell{Mean of Reference\\Population} & \makecell{Mean of Control\\Group (no rule)} & \makecell{Mean of Treated\\Group (has rule)} & SMD for Control & SMD for Treated \\
\midrule
Community Size & 1661.26 & 1328.24 & 2373.23 & -0.02 & 0.00 \\
Discussion & 0.11 & 0.10 & 0.14 & 0.00 & 0.00 \\
Hobby & 0.51 & 0.50 & 0.53 & 0.00 & 0.00 \\
Identity & 0.17 & 0.17 & 0.17 & 0.00 & -0.01 \\
Media & 0.14 & 0.16 & 0.10 & 0.00 & 0.01 \\
Memes & 0.05 & 0.05 & 0.04 & 0.00 & 0.00 \\
News & 0.02 & 0.03 & 0.02 & 0.01 & 0.00 \\
\bottomrule
\end{tabular}
\end{table}

\subsection{Covariate Balance for Links \& External Content Rules}
\begin{table}[H]
\centering
\label{tab:balance_links_external_content}
\begin{tabular}{lccccc}
\toprule
Covariate & \makecell{Mean of Reference\\Population} & \makecell{Mean of Control\\Group (no rule)} & \makecell{Mean of Treated\\Group (has rule)} & SMD for Control & SMD for Treated \\
\midrule
Community Size & 1661.26 & 1414.87 & 2026.90 & 0.21 & 0.01 \\
Discussion & 0.11 & 0.13 & 0.08 & 0.03 & 0.00 \\
Hobby & 0.51 & 0.50 & 0.52 & -0.01 & 0.01 \\
Identity & 0.17 & 0.17 & 0.17 & 0.00 & -0.01 \\
Media & 0.14 & 0.14 & 0.14 & -0.01 & 0.00 \\
Memes & 0.05 & 0.05 & 0.05 & 0.00 & -0.01 \\
News & 0.02 & 0.02 & 0.04 & 0.00 & 0.00 \\
\bottomrule
\end{tabular}
\end{table}

\subsection{Covariate Balance for Images Rules}
\begin{table}[H]
\centering
\label{tab:balance_images}
\begin{tabular}{lccccc}
\toprule
Covariate & \makecell{Mean of Reference\\Population} & \makecell{Mean of Control\\Group (no rule)} & \makecell{Mean of Treated\\Group (has rule)} & SMD for Control & SMD for Treated \\
\midrule
Community Size & 1661.26 & 1043.11 & 3110.61 & 0.19 & 0.01 \\
Discussion & 0.11 & 0.14 & 0.05 & -0.01 & 0.04 \\
Hobby & 0.51 & 0.52 & 0.49 & -0.02 & -0.01 \\
Identity & 0.17 & 0.19 & 0.11 & -0.01 & -0.01 \\
Media & 0.14 & 0.10 & 0.22 & -0.01 & 0.00 \\
Memes & 0.05 & 0.03 & 0.10 & -0.03 & 0.00 \\
News & 0.02 & 0.02 & 0.03 & 0.17 & -0.01 \\
\bottomrule
\end{tabular}
\end{table}

\subsection{Covariate Balance for Commercialization Rules}
\begin{table}[H]
\centering
\label{tab:balance_commercialization}
\begin{tabular}{lccccc}
\toprule
Covariate & \makecell{Mean of Reference\\Population} & \makecell{Mean of Control\\Group (no rule)} & \makecell{Mean of Treated\\Group (has rule)} & SMD for Control & SMD for Treated \\
\midrule
Community Size & 1661.26 & 1384.51 & 1954.49 & -0.01 & 0.00 \\
Discussion & 0.11 & 0.15 & 0.08 & 0.00 & 0.00 \\
Hobby & 0.51 & 0.38 & 0.65 & 0.00 & 0.01 \\
Identity & 0.17 & 0.16 & 0.18 & 0.00 & 0.01 \\
Media & 0.14 & 0.20 & 0.07 & 0.00 & 0.01 \\
Memes & 0.05 & 0.08 & 0.01 & 0.01 & -0.05 \\
News & 0.02 & 0.03 & 0.01 & 0.00 & 0.00 \\
\bottomrule
\end{tabular}
\end{table}

\subsection{Covariate Balance for Illegal Content Rules}
\begin{table}[H]
\centering
\label{tab:balance_illegal_content}
\begin{tabular}{lccccc}
\toprule
Covariate & \makecell{Mean of Reference\\Population} & \makecell{Mean of Control\\Group (no rule)} & \makecell{Mean of Treated\\Group (has rule)} & SMD for Control & SMD for Treated \\
\midrule
Community Size & 1661.26 & 1596.03 & 1856.06 & 0.02 & 0.02 \\
Discussion & 0.11 & 0.12 & 0.10 & 0.00 & 0.03 \\
Hobby & 0.51 & 0.48 & 0.59 & 0.00 & -0.01 \\
Identity & 0.17 & 0.17 & 0.17 & 0.00 & 0.01 \\
Media & 0.14 & 0.15 & 0.11 & 0.00 & 0.02 \\
Memes & 0.05 & 0.06 & 0.03 & 0.01 & -0.04 \\
News & 0.02 & 0.03 & 0.01 & 0.01 & -0.05 \\
\bottomrule
\end{tabular}
\end{table}

\subsection{Covariate Balance for Divisive Content Rules}
\begin{table}[H]
\centering
\label{tab:balance_divisive_content}
\begin{tabular}{lccccc}
\toprule
Covariate & \makecell{Mean of Reference\\Population} & \makecell{Mean of Control\\Group (no rule)} & \makecell{Mean of Treated\\Group (has rule)} & SMD for Control & SMD for Treated \\
\midrule
Community Size & 1661.26 & 1316.22 & 2772.01 & \textbf{0.96} & 0.04 \\
Discussion & 0.11 & 0.10 & 0.16 & 0.09 & 0.01 \\
Hobby & 0.51 & 0.56 & 0.34 & -0.03 & 0.01 \\
Identity & 0.17 & 0.14 & 0.25 & -0.01 & -0.01 \\
Media & 0.14 & 0.14 & 0.12 & -0.01 & -0.02 \\
Memes & 0.05 & 0.04 & 0.08 & -0.02 & 0.01 \\
News & 0.02 & 0.02 & 0.04 & -0.02 & 0.03 \\
\bottomrule
\end{tabular}
\end{table}

\subsection{Covariate Balance for Respect for Others Rules}
\begin{table}[H]
\centering
\label{tab:balance_respect_for_others}
\begin{tabular}{lccccc}
\toprule
Covariate & \makecell{Mean of Reference\\Population} & \makecell{Mean of Control\\Group (no rule)} & \makecell{Mean of Treated\\Group (has rule)} & SMD for Control & SMD for Treated \\
\midrule
Community Size & 1661.26 & 824.14 & 1934.97 & \textbf{0.28} & 0.00 \\
Discussion & 0.11 & 0.10 & 0.12 & 0.15 & 0.00 \\
Hobby & 0.51 & 0.53 & 0.50 & -0.07 & 0.01 \\
Identity & 0.17 & 0.10 & 0.19 & -0.06 & 0.00 \\
Media & 0.14 & 0.18 & 0.12 & 0.01 & -0.02 \\
Memes & 0.05 & 0.06 & 0.05 & 0.03 & -0.01 \\
News & 0.02 & 0.02 & 0.02 & -0.02 & 0.00 \\
\bottomrule
\end{tabular}
\end{table}

\subsection{Covariate Balance for Brigading Rules}
\begin{table}[H]
\centering
\label{tab:balance_brigading}
\begin{tabular}{lccccc}
\toprule
Covariate & \makecell{Mean of Reference\\Population} & \makecell{Mean of Control\\Group (no rule)} & \makecell{Mean of Treated\\Group (has rule)} & SMD for Control & SMD for Treated \\
\midrule
Community Size & 1661.26 & 1461.99 & 3498.79 & 0.07 & 0.11 \\
Discussion & 0.11 & 0.11 & 0.16 & 0.01 & 0.02 \\
Hobby & 0.51 & 0.52 & 0.43 & 0.00 & -0.01 \\
Identity & 0.17 & 0.16 & 0.20 & 0.00 & -0.01 \\
Media & 0.14 & 0.14 & 0.08 & 0.00 & 0.01 \\
Memes & 0.05 & 0.04 & 0.10 & -0.01 & 0.00 \\
News & 0.02 & 0.02 & 0.02 & 0.00 & -0.01 \\
\bottomrule
\end{tabular}
\end{table}

\subsection{Covariate Balance for Ban Mentioned Rules}
\begin{table}[H]
\centering
\label{tab:balance_ban_mentioned}
\begin{tabular}{lccccc}
\toprule
Covariate & \makecell{Mean of Reference\\Population} & \makecell{Mean of Control\\Group (no rule)} & \makecell{Mean of Treated\\Group (has rule)} & SMD for Control & SMD for Treated \\
\midrule
Community Size & 1661.26 & 1556.12 & 3432.80 & 0.01 & 0.08 \\
Discussion & 0.11 & 0.11 & 0.16 & 0.00 & 0.01 \\
Hobby & 0.51 & 0.51 & 0.43 & 0.00 & 0.05 \\
Identity & 0.17 & 0.17 & 0.15 & 0.00 & -0.04 \\
Media & 0.14 & 0.14 & 0.15 & 0.00 & -0.02 \\
Memes & 0.05 & 0.05 & 0.08 & 0.00 & -0.02 \\
News & 0.02 & 0.02 & 0.03 & 0.00 & -0.01 \\
\bottomrule
\end{tabular}
\end{table}

\subsection{Covariate Balance for Karma/Score Mentioned Rules}
\begin{table}[H]
\centering
\label{tab:balance_karma_score_mentioned}
\begin{tabular}{lccccc}
\toprule
Covariate & \makecell{Mean of Reference\\Population} & \makecell{Mean of Control\\Group (no rule)} & \makecell{Mean of Treated\\Group (has rule)} & SMD for Control & SMD for Treated \\
\midrule
Community Size & 1661.26 & 1519.41 & 4628.85 & 0.00 & 0.03 \\
Discussion & 0.11 & 0.11 & 0.13 & 0.00 & 0.05 \\
Hobby & 0.51 & 0.51 & 0.45 & 0.00 & 0.01 \\
Identity & 0.17 & 0.17 & 0.06 & 0.00 & -0.06 \\
Media & 0.14 & 0.14 & 0.14 & 0.00 & 0.00 \\
Memes & 0.05 & 0.04 & 0.20 & -0.01 & 0.02 \\
News & 0.02 & 0.02 & 0.02 & 0.00 & -0.01 \\
\bottomrule
\end{tabular}
\end{table}

\subsection{Dataset Datasheet}\label{app:dataset_ethics}
\xhdr{Motivation}
These data were curated with the goal of understanding the evolution of governance across Reddit by researchers at the University of Washington. This work was was supported by the Office of Naval Research (\#N00014-21-1-2154), NSF grant IIS-1901386, NSF CAREER IIS-2142794, NSF grant CNS-2025022, and the Bill \& Melinda Gates Foundation (INV-004841).

\xhdr{Composition}

Our published dataset consists of rules extracted by regex from at-most-weekly snapshots provided by the Wayback Machine, an archival service offered by the nonprofit Internet Archive. The instances composing the scraped rules consist of the date of the snapshot, the community name, and all rules found within the snapshot. In total, these instances span 2,472,877 rule records (73,087 unique rules) across 6,120 communities from April 2018 to December 2023. Some communities could not be scraped due to privacy settings, lack of rules within sidebar text, or non-standard formatting. Snapshots without found rules text are not included in the dataset. Data collection is further described in \sect\ref{sec:data}.

The labeled dataset was created using a language model to classify scraped rules. Rules with an edit distance less than 3 characters, along with rules that differ only in punctuation and casing, are assigned the same labels. These data consist of records identical to the ones above, with additional boolean data for each rule type discussed in our codebook (\sect\ref{sec:codebook}). Additional gaps exist for rules filtered by OpenAI's content policy. Details of model performance are included in \tab\ref{tab:model_performance}. The datasets are self contained. The datasets do not contain confidential or sensitive information such as usernames. 

\xhdr{Collection Process}

The dataset was created by applying a labeling pipeline to rule text scraped from the sidebar of snapshots from old.reddit.com, curated by Internet Archive's Wayback Machine. This is described in detail in \sect\ref{sec:codebook}. No sampling was performed in the production of the dataset, and no laborers were used for labeling tasks beyond the authors' labeling of training and test sets. These data span 6,120 communities from  April 2018 to December 2023. This work was approved by the University of Washington IRB under ID number STUDY00011457.

\xhdr{Preprocessing/Cleaning/Labeling}
Several steps of preprocessing were applied to the raw dataset to reduce the cost of labeling, as described in detail in \sect\ref{sec:codebook}. We make both the original scraped data public alongside the results of our labeling pipeline. Additionally, we make all code used for preprocessing, clustering, and labeling public.

\xhdr{Uses}
Our datasets have already been used for the analyses of rules, changes, and impacts thereof, as presented in this paper. A link to this paper will be clearly posted along with the datasets. These datasets can be used for many other analyses of moderation and governance on Reddit. We do not believe there are any tasks for which the dataset could reasonably be used that would result in breaches of users privacy or other harms.

\xhdr{Distribution}
Our datasets will be distributed publicly upon peer-reviewed publication of this paper. They will be published with an open and clearly stated data usage license complaint with Reddit and Internet Archive's data usage and redistribution policies. The datasets have no IP restrictions on them, and, as far as we are aware, are not subject to export controls or regulatory restrictions.

\xhdr{Maintenance}
Our dataset will be hosted publicly on our University-hosted webpage, as well as on a public academic dataset hosting service to be determined at the time of publication. Contact information for the dataset (the corresponding author of this paper) will be made clear and public. Our datasets are derived only from publicly-posted data which we have processed, and we will provide a channel for users and communities to have their data removed from our dataset. 

\end{document}